\documentclass[pra,aps,notitlepage,amsmath,amssymb,superscriptaddress,10pt]{revtex4-2}

\usepackage[dvipdfmx]{graphicx}
\usepackage{amsmath,amssymb,amsthm,bm}
\usepackage{xcolor}
\newtheorem{Lem}{Lemma}

\newcommand{\sqb}[1]{\displaystyle \left[ #1 \right]}
\newcommand{\rb}[1]{\displaystyle \left( #1 \right)}
\newcommand{\cab}[1]{\displaystyle \left| #1 \right|}

\newcommand{\outerb}[2]{\displaystyle \ket{ #1 } \hspace{-2pt}\bra{ #2 }}

\newcommand{\nm}[1]{\displaystyle \left\Vert {#1} \right\Vert }
\newcommand{\tnm}[1]{\displaystyle \left\Vert {#1} \right\Vert _\mathrm{tr}}
\newcommand{\dnm}[1]{\displaystyle \left\Vert {#1} \right\Vert _\diamond}
\def\qed{\hfill $\Box$}

\newcommand{\bra}[1]{\langle {#1} \vert}

\newcommand{\ket}[1]{\vert {#1} \rangle}

\newcommand{\braket}[2]{\langle {#1} \vert {#2} \rangle}

\newcommand{\abs}[1]{\left| {#1} \right|}
\newcommand{\norm}[1]{\abs{\abs{#1}}}

\newcommand{\Tr}[0]{{\mathrm{Tr}}}

\begin{document}

\title{Transferring quantum information between a quantum system with limited control and a quantum computer}
\author{Ryosuke Sakai}
 \affiliation{Department of Physics, Graduate School of Science, The University of Tokyo, Tokyo, Japan} 
\author{Akihito Soeda}
\affiliation{Department of Physics, Graduate School of Science, The University of Tokyo, Tokyo, Japan} 
\author{Mio Murao}
\affiliation{Department of Physics, Graduate School of Science, The University of Tokyo, Tokyo, Japan}
\affiliation{Trans-scale Quantum Science Institute, The University of Tokyo, Bunkyo-ku, Tokyo 113-0033, Japan} 

\begin{abstract}
We consider a hybrid quantum system consisting of a qubit system continuously evolving according to its fixed own Hamiltonian and a quantum computer. The qubit system couples to a quantum computer through a fixed interaction Hamiltonian, which can only be switched on and off. We present quantum algorithms to approximately transfer quantum information between the qubit system with limited control and the quantum computer under this setting. Our algorithms are programmed by the gate sequences in a closed formula for a given interface interaction Hamiltonian.
\end{abstract}

\maketitle

\section{Introduction} \label{sec:intro}

Quantum information processing requires high controllability to prepare and manipulate quantum systems.  However, actual physical systems described by Hamiltonians are not necessarily fully controllable due to the limitations for controllable parameters in the system, even for the cases where dissipation and decoherence are negligible.  In spite of the difficulties, constructions of small scale highly-controllable quantum systems have been demonstrated in various quantum systems such as superconducting systems, quantum optical systems, trapped ion systems and NMR systems, where we can apply a wide range of quantum manipulations. However, large-scale realizations of such highly-controllable quantum systems are still under development.    Instead of requiring every part of the quantum system to be highly-controllable, one possible way to achieve larger scale highly-controllable quantum information processing is to extend a small scale highly-controllable quantum system by combining with a physical system with limited controllability and introducing algorithms for controlling the physical system.

Seeking methods to manipulate a quantum system with limited controllability by combining with a controllable quantum system has been studied as the context of \emph{quantum control theory of dynamics} \cite{SL:CQF, A:IQC, DP:QT, AT:MC, OM:PU, SL:CQI, DB:LocalCon, RH:Localq, RH:Controlq, DB:FullControl, DB:Protocol, DV:defrag, DB:Scalableq}, which aims to implement desired operations by adding extra Hamiltonians with controllable parameters on the original Hamiltonian of the uncontrollable system.    In the quantum control theory of dynamics, the total Hamiltonian is typically given by $H(t) = H_S + \sum _{k=1}^L \alpha _k (t) H_k$ where $H_S$ is the system Hamiltonian without controllable parameters and $\{ H_k \}$ are the additional control Hamiltonians that their contributions can be independently tuned by adjusting a time dependent parameter $\alpha_k(t)$.  It is often assumed that $H_S$ is the Hamiltonian of a composite system which includes interactions among the subsystems and $H_k$'s are local Hamiltonians on the subsystems \cite{OM:PU, SL:CQI, DB:LocalCon, RH:Localq, RH:Controlq, DB:Scalableq}.    For the systems such as NMR systems, the time scale of the control Hamiltonian dynamics can be much faster than the uncontrollable system dynamics.  That is, the control Hamiltonians can be applied as intensive pulses for such systems and the methods for the time optimal control with intensive pulses have been developed in \cite{OM:PU, SL:CQI, DB:LocalCon, RH:Localq, RH:Controlq, DB:Scalableq}.  In many other quantum systems, however,  the range of the absolute value of $\alpha_k(t)$ is limited and thus it is not possible to straightforwardly apply the methods using intensive pulses.

We consider to combine a physical system with limited controllability with a quantum computer instead of Hamiltonian dynamics.  Quantum computers are idealized systems, i.e., we can perform arbitrary operations allowed in quantum mechanics, which can be represented by the quantum circuit model.  In the preceding works on achieving full control by locally induced relaxation \cite{DB:Scalableq, DB:FullControl, DB:Protocol, DV:defrag}, the authors also considered a physical system without controllable parameters coupling to a highly controllable system that can be regarded as a quantum computer.  They have shown that in principle, any state of a physical system can be transferred to the quantum computer and can be transferred back from the quantum computer to the physical system, if the system Hamiltonian $H_S$ and the fixed interaction Hamiltonian $H_\mathrm{int}$ describing the coupling satisfy certain conditions that can be interpreted to induce relaxation of the physical system.   Since any transformations can be applied in a quantum computer, ability of this type of \emph{input-output (IO) operations}, namely, the ability to transfer quantum states from the physical system to the quantum computer (an output process) and transferring back quantum states from the quantum computer to the physical system (an input process) achieves the full dynamical control of the physical system via the quantum computer.

Algorithms achieving \emph{approximate} IO operations between a physical system without controllability composed of several subsystems and a quantum computer coupled for a fixed interaction Hamiltonian $H_\mathrm{int}$ between the physical system and a part of the quantum computers are also presented in \cite{DB:FullControl, DB:Protocol, DV:defrag}.  These algorithms assume that the time scale of performing operations on the quantum computer is much faster than the dynamics generated by $H_S + H_\mathrm{int}$.  However, the time cost for performing operations on the quantum computer is not necessary negligible in general, and the algorithms taking the effect of the time cost are desired for physical implementations.

In this preprint, we investigate algorithms to achieve the approximate IO operations for a \emph{single-qubit} physical system even when the time cost on running algorithms in the quantum computer is not negligible.   To obtain such algorithms, we relax the requirement for the controllability of the coupling between the physical system and the quantum computer.    Namely, we assume that we can switch $H_\mathrm{int}$ on and off.   That is, the physical system evolves either by $H_S$ or $H_S + H_\mathrm{int}$ whereas we cannot control the choice of $H_\mathrm{int}$.   Note that this assumption does not provide much power on the controllability of the physical system comparing to the situations considered in \cite{DB:FullControl, DB:Protocol, DV:defrag}, since assuming the infinitesimal time cost for operations in a quantum computer guarantees to be able to neglect the effect of $H_S$ and $H_\mathrm{int}$ while the operations of quantum computers are applied.

We present two concrete algorithms implementing the output process for the approximate IO operations represented by quantum circuits. The first algorithm requires the quantum computer to have a linearly-growing quantum memory, which is referred as \emph{the linearly growing size (LS) algorithm}.   The second algorithm is an improved algorithm that requires only a constant-size quantum memory, which is referred as \emph{the constant size (CS) algorithm}.  The quantum circuits representing both of the two algorithms contain the interface unitary $U_\mathrm{int}$ generated by $H_S + H_\mathrm{int}$ satisfying certain conditions. 
We derive the conditions for  $U_\mathrm{int}$ for implementing the approximate IO operations by using each algorithm.   We also evaluate the accuracy of the implemented operations in the diamond norm \cite{diamondnorm}, and show that the CS algorithm achieves the IO operation with an arbitrary accuracy by using a finite-size quantum memory. We present the algorithms implementing the input process corresponding to the LS and CS algorithms for the approximate IO operations.

This preprint is organized as follows. We present our settings and clarify the assumptions on the operations on a physical system and quantum computers in Sec.~\ref{sec:settings}.  We show a lemma which evaluates the accuracy of the desired operation for the physical system by giving trial approximate IO operations in the diamond norm in Sec.~\ref{sec:acc}.  In Sec.~\ref{sec:LS} and \ref{sec:CS}, we show the LS/CS algorithms programmed by the gate sequences for the trial approximate output operation, and demonstrate the necessary conditions for interface unitary operations.  We complete the algorithms for the IO operation by applying a final adjustment for the LS/CS algorithms in Sec.~\ref{sec:rest}.    We will show how to prepare the interface unitary operations satisfying such conditions by the given interaction Hamiltonian, and discuss whether two specific systems can be manipulatable in Sec.~\ref{sec:Ex}.    A summary of our results is presented in Sec.~\ref{sec:conc}.  

This preprint is based on the master thesis of Ryosuke Sakai \cite{SakaiMasterThesis} submitted to Department of Physics, Graduate School of Science, The University of Tokyo in 2016.  The content of this preprint is submitted to a chapter (Chapter 14) in the book, {\it Hybrid Quantum Systems}, edited by Yoshiro Hirayama, Koji Ishibashi, and Kae Nemoto, and published by Springer-Nature Singapore Pte Ltd..

\section{Settings}\label{sec:settings}

\subsection{Assumptions for a physical system coupling to a quantum computer} \label{sec:allowed}

We consider a physical system denoted by $S$ represented by a two-dimensional Hilbert space $\mathcal{H}_S$, a qubit system.  System $S$ evolves according to a fixed time-independent self-Hamiltonian $H_S$ and has limited controllability.  Our goal is to apply arbitrary quantum operations on $S$ by coupling to a quantum computer.    We assume that the quantum computer in contrast to $S$ is able to perform arbitrary quantum operations represented by quantum circuits.  We set the Planck units to be $\hbar = 1$.  For simplicity, we sometimes abbreviate the the joint system of $S$ and $I$ by $SI$, the joint system of $I$ and $R$ by $IR$ and the total system of $S$, $I$, and $R$ by $SIR$.   We refer to a basis $\{\ket{\Psi _j}_{SI}\}_j$ of the joint system $SI$ as a \emph{local basis} when each element of the basis states can be represented as a product of normalized pure states. 

We model the coupling between the system $S$ and the quantum computer by introducing two subsystems in the quantum computer, a register system $R$ and an interface system $I$.   Subsystem $R$ is the main processing part of the quantum computer consisting of $N$ qubits represented by a $2^N$-dimensional Hilbert space $\mathcal{H}_R^N$.
The computational basis of each qubit in $R$ is represented by $\{ \ket{0}, \ket{1} \}$.   We assume that any quantum operations can be freely performed on $R$.  Subsystem $I$ is a mediating system in the quantum computer which directly couples to $S$ by a \emph{fixed} interaction between $S$ and $I$.   We set $I$ to be a single qubit system represented by a two-dimensional Hilbert space $\mathcal{H}_I$.  We assume that we can only control switching on and off of a single fixed interaction Hamiltonian represented by $H_\mathrm{int}$ on $\mathcal{H}_S \otimes \mathcal{H}_I$.  The total Hamiltonian of system $SI$ is given by
\begin{align}
H_{SI}(t) := H_S \otimes I_I + \alpha _\mathrm{on-off}(t) H_\mathrm{int}, \label{eq:Ht}
\end{align}
where $\alpha _\mathrm{on-off}(t)$ is a \emph{binary} switching parameter for time $t$ that can be designed arbitrarily.  We assume that $I$ behaves as a part of the quantum computer, i.e., any quantum operations can be performed on $\mathcal{H}_I \otimes \mathcal{H}_R^N$ when $\alpha _\mathrm{on-off}(t)=0$, but it is dominated by the Hamiltonian dynamics according to the Hamiltonian on $\mathcal{H}_S \otimes \mathcal{H}_I$ given by
\begin{align}
H_{\mathrm{on}}:= H_S \otimes I_I + H_\mathrm{int}, \label{eq:Hon}
\end{align}
when $H_\mathrm{int}$ is switched on ($\alpha _\mathrm{on-off}(t) = 1$), in which case all additional operations on $I$ are prohibited.

\begin{figure}[th]
\centering
\includegraphics[width=120truemm]{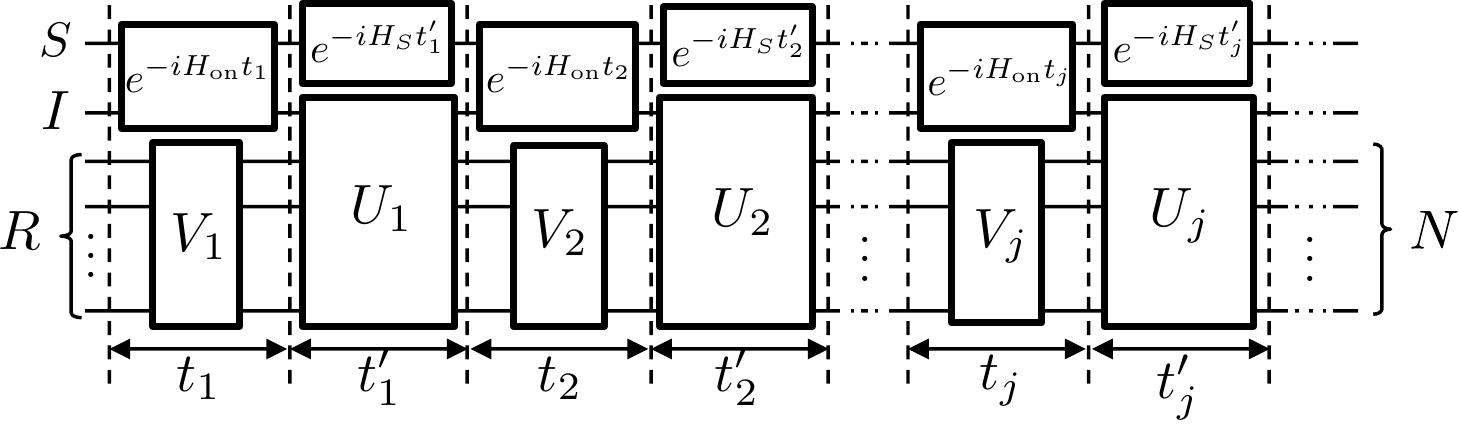}
\caption{A quantum circuit representation of a set of most general unitary operators $\mathbb{LU}_{H_S,H_\mathrm{int}}^N$ allowed in our setting. $t_j$ and $t'_j$ (for $j = 1,2,\dots$) are duration time of switching on and off, respectively. $V_j$ and $U_j$ are arbitrary unitary operators on $\mathcal{H}_{R}^N$ and $\mathcal{H}_{I} \otimes \mathcal{H}_{R}^N$, respectively.
}
\label{fig:LUH}
\end{figure}

The quantum circuit represented by Fig.~\ref{fig:LUH} covers all possible deterministic operations under the assumptions made for systems $S$, $I$, and $R$. We refer to the set of unitary operators on $\mathcal{H}_S \otimes \mathcal{H}_I \otimes \mathcal{H}_R^N$ of which quantum circuit decompositions are represented by Fig.~\ref{fig:LUH} as $\mathbb{LU}^N_{H_S, H_\mathrm{int}}$.  

\begin{figure}[th]
\centering
\includegraphics[width=120truemm]{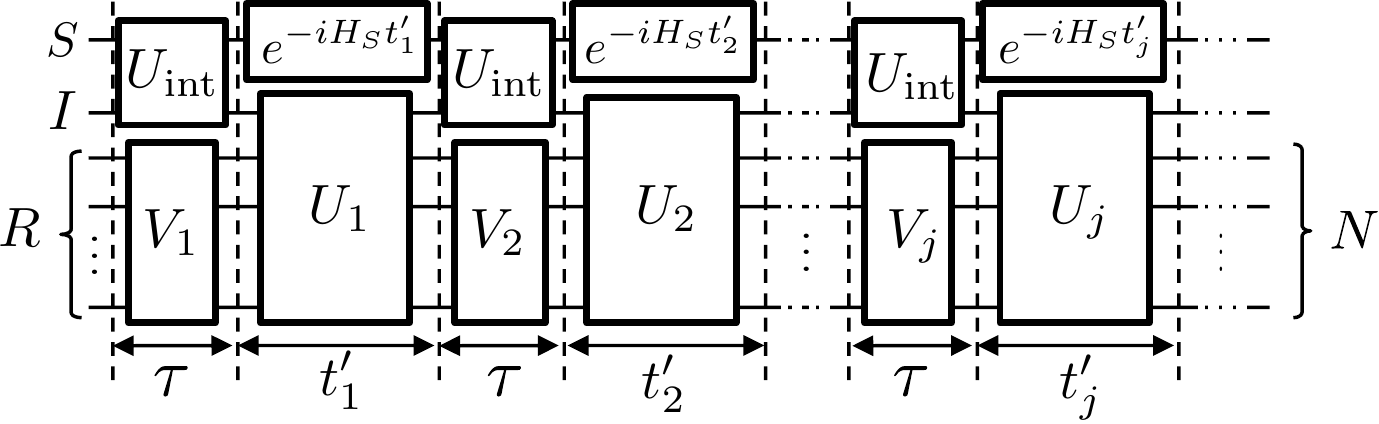}
\caption{A quantum circuit representation of a set of unitary operators $\mathbb{LU}_{H_S, U_\mathrm{int}}^N$ in the simplified setting where the interaction Hamiltonian is switched on for a fixed duration time $\tau$.  $V_j$ and $U_j$ are arbitrary unitary operators on $\mathcal{H}_{R}^N$ and $\mathcal{H}_{I} \otimes \mathcal{H}_{R}^N$, respectively.}
\label{fig:LUUH}
\end{figure}

The independent parameters $\{ t_i \}$ and $\{ t'_j \}$ in the quantum circuit representation of $\mathbb{LU}_{H_S,H_\mathrm{int}}^N$ introduce too large degrees of freedom to handle concrete constructions of the approximate IO operations.  To handle this difficulty, we introduce a simplified setting by choosing a fixed time duration $\tau = t_1 = t_2 = \cdots$ on the switched on time of  $H_\mathrm{int}$.  This simplification is equivalent to applying a fixed unitary operator
\begin{align}
U_\mathrm{int}= e^{-i H_{\mathrm{on}}\tau} \label{eq:timeev}
\end{align}
on $\mathcal{H}_S \otimes \mathcal{H}_I$ while $H_\mathrm{int}$ is switched on.  We refer to $U_\mathrm{int}$ as an \emph{interface unitary operator}.   The quantum circuit represented by Fig.~\ref{fig:LUUH} covers all possible unitary operators in this simplified setting.    We refer to the set of unitary operators on $\mathcal{H}_S \otimes \mathcal{H}_I \otimes \mathcal{H}_R^N$ represented by Fig.~\ref{fig:LUUH} as $\mathbb{LU}_{H_S, U_\mathrm{int}}^N$ which is obviously a subset of $\mathbb{LU}_{H_S, H_\mathrm{int}}^N$.  We construct the gate sequences of approximated IO algorithms in this setting for a given interface unitary operator $U_\mathrm{int}$ in $\mathbb{LU}_{H_S, U_\mathrm{int}}^N$.  

\begin{figure}[th]
\centering
\includegraphics[width=90truemm]{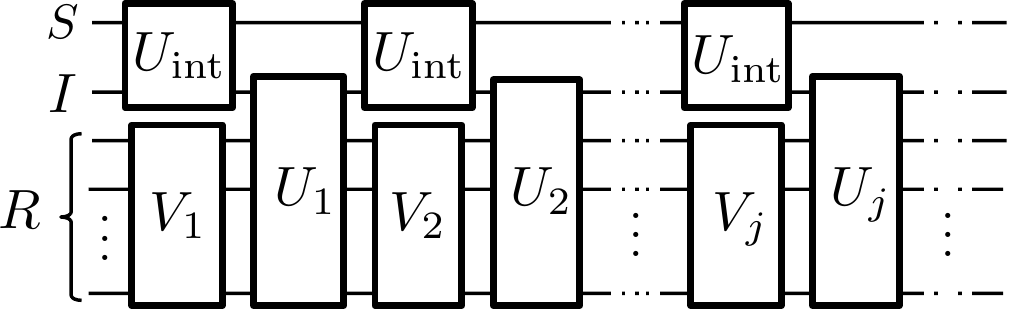}
\caption{A quantum circuit representation of a set of unitary operators $\mathbb{LU}_{U_\mathrm{int}}^N$.  $V_j$ and $U_j$ are arbitrary unitary operators on $\mathcal{H}_{R}$ and $\mathcal{H}_{I} \otimes \mathcal{H}_{R}$, respectively.
}
\label{fig:LUU}
\end{figure}

We also introduce a further simplified setting where we can switch off $H_S$ while any $U_j$ is applied on system $IR$ represented by the quantum circuit given by Fig.~\ref{fig:LUU}.  In this case, the parameters $\{ t'_j\}$ can be also eliminated. We refer to the set of unitary operators on $\mathcal{H}_S \otimes \mathcal{H}_I \otimes \mathcal{H}_R^N$ represented by Fig.~\ref{fig:LUU} as $\mathbb{LU}_{U_\mathrm{int}}$.  As the first step, we present algorithms for the approximate IO operations in this settings in Sec.~\ref{sec:LS} and Sec.~\ref{sec:CS}.    One of the motivations of our work is to consider the non-negligible time cost for performing operations on the quantum computer, therefore this simplified setting may look similar to the case where the time evolution due to $H_S$ during the time operating quantum computer is ignored.  However, we show that both of the proposed algorithms in the setting for $\mathbb{LU}_{U_\mathrm{int}}$ can be extended to the setting for $\mathbb{LU}_{H_S, U_\mathrm{int}}^N$, which fully includes the time cost of quantum operation in the quantum computer by introducing a final adjustment utilizing the parameters $\{ t'_j \}$ in Sec.~\ref{sec:rest}.

There are still too many parameters in a whole set of two qubit unitary operators $U_\mathrm{int}$ to construct concrete algorithms.  We restrict our consideration of $U_\mathrm{int}$ to the ones of which matrix representation in a local basis $\{ \ket{i}_S \otimes \ket{j}_I \}_{i,j=0,1}$ of $\mathcal{H}_S \otimes \mathcal{H}_I$ is given by 
\begin{align}
\big\{ \bra{i}_S \bra{j}_I U_\mathrm{int} \ket{k}_S \ket{l}_I \big\}_{i,j,k,l=0,1}
= \begin{bmatrix}
u_{00,00} & u_{00,01} & u_{00,10} & u_{00,11} \\
u_{01,00} & u_{01,01} & u_{01,10} & u_{01,11} \\
0 & u_{10,01} & u_{10,10} & u_{10,11} \\
0 & u_{11,01} & u_{11,10} & u_{11,11} \\
\end{bmatrix}, \label{eq:matU} 
\end{align}
where $u_{ij,kl} (i,j,k,l = 0,1)$ are complex numbers.  We denote a set of such unitary operators on $\mathcal{H}_S \otimes \mathcal{H}_I$ by $\mathbb{U}^\star$.  We will show that for all $U_\mathrm{int} \in \mathbb{U}^\star$ there exist approximated output processes of the IO operations by presenting the LS and CS algorithm in Sec.~\ref{sec:LS} and Sec.~\ref{sec:CS}.   Thus we call the local basis $\{ \ket{i}_S \otimes \ket{j}_I \}_{i,j=0,1}$ satisfying Eq.~(\ref{eq:matU}) for a given interface unitary operator $U_\mathrm{int}$ as the \emph{feasible basis} of $U_\mathrm{int}$.   We assume that the computational bases $\{ \ket{0},\ket{1} \}$ of $SI$ are chosen to be the basis of each system for the feasible basis in the following. 

\section{Approximate IO operations evaluated in terms of the diamond norm} \label{sec:acc}

In our algorithms, we consider a procedure to implement an arbitrary quantum operation denoted by $\mathcal{M}$ on $S$ by the input-output (IO) operations consisting of the following three steps.  
\begin{enumerate}
\item{The {\it output process} from $S$ to $I$:  Design an operation in $\mathbb{LU}_{H_S,U_\mathrm{int}}^N$ to transfer an arbitrary state from $S$ to $I$.}
\item{Application of $\mathcal{M}$ in $I$:  Apply a desired operation $\mathcal{M}$ on $I$, which is always possible while $H_{int}$ is switched off in our definition. }
\item{The {\it input process} from $I$ to $S$: Design an operation in $\mathbb{LU}_{H_S,U_\mathrm{int}}^N$ to transfer back the state from $I$ to $S$.}
\end{enumerate}

Obviously,  if a unitary operator called a {\it half-swap} operation $U_\mathrm{hswap}$ on $\mathcal{H}_S \otimes \mathcal{H}_I$ satisfying $U_\mathrm{hswap}\ket{0}_S \ket{0}_I =\ket{0}_S \ket{0}_I$ and $U_\mathrm{hswap} \ket{1}_S \ket{0}_I =\ket{0}_S \ket{1}_I$ is in $\mathbb{LU}_{H_S,U_\mathrm{int}}^N$, the output process is \emph{exactly} achievable by setting an initial state of $I$ to be $\ket{0}$ and applying $U_\mathrm{hswap}$.   We consider the cases where such a convenient unitary is \emph{not} available in $\mathbb{LU}_{H_S,U_\mathrm{int}}^N$ and investigate a strategy for \emph{approximately} implementing the output and input processes by applying unitary operations on $IR$.

In the following of this section, we rename a part of $R$ used to implement $\mathcal{M}$ by a unitary operation on $I$ and the extra ancilla system as system $E$, and remove $E$ from $R$.   We set that $R$ consists of $M$ qubits and $E$ consists of $N-M$ qubits.  We also set the initial state of $IR$ to be $\ket{0}_I \otimes \ket{0}_R^{\otimes M}$.  For approximation, we evaluate the difference between the desired operation $\mathcal{M}$ and the approximated IO operation in terms of the diamond norm to guarantee the worst case to be within a preset bound.

Our strategy is to consider a one-parameter family of unitary operations $T_M^\mathrm{out}(\xi _\mathrm{out})$ on $\mathcal{H}_S \otimes \mathcal{H}_I \otimes \mathcal{H}_R^M$ for approximately implementing the output process by transforming an arbitrary state $\ket{\psi_S}_S := a_S \ket{0}_S + b_S \ket{1}_S 
$ of $S$ satisfying $| a_S|^2 + | b_S |^2 =1$ for $a_S, b_S \in \mathbb{C}$ and $\ket{0}_I \otimes \ket{0}_R^{\otimes M}$ of $IR$ as
\begin{align}
T_M^\mathrm{out}(\xi _\mathrm{out}) \ket{\psi _S} _S \ket{0}_I \ket{0}^{\otimes M}_R 
= \ket{0}_S \otimes \big( a_S \ket{0}_I + b_S \sqrt{1-\xi _\mathrm{out}^2} \ket{1}_I \big) \otimes \ket{0}_R^{\otimes M} 
+ b_S \xi _\mathrm{out} \ket{1}_S \ket{g_\mathrm{out}}_{IR},\label{eq:ToutM}
\end{align}
where the parameter $\xi _\mathrm{out}$ is set to be $0 \leq \xi _\mathrm{out} \leq 1$ and $\ket{g_\mathrm{out}}_{IR}$ can be any state.   The action of $T_M^\mathrm{out}(0)$ on $\ket{\psi_S}_S \otimes \ket{0}_I \otimes \ket{0}_R^{\otimes M}$ achieves the exact output process from $S$ to $I$. 

Similarly, we consider another one-parameter family of unitary operation $T_M^\mathrm{in}(\xi _\mathrm{in})$ on on $\mathcal{H}_S \otimes \mathcal{H}_I \otimes \mathcal{H}_R^M$
for approximately implementing the input process satisfying, namely,  
 \begin{align}
\big( T_M^\mathrm{in} (\xi _\mathrm{in}) \big) ^\dagger \ket{\psi _S} _S \ket{0}_I \ket{0}^{\otimes M}_R 
= \ket{0}_S \otimes \big( a_S \ket{0}_I + b_S \sqrt{1-\xi _\mathrm{in}^2} \ket{1}_I \big) \otimes \ket{0}_R^{\otimes M} 
+ b_S \xi _\mathrm{in} \ket{1}_S \ket{g_\mathrm{in}}, \label{eq:Tin}
\end{align}
where $0 \leq \xi _\mathrm{in} \leq 1$ and $\ket{g_\mathrm{in}} \in \mathbb{S}( \mathcal{H}_I \otimes \mathcal{H}_R^M)$ can be any state.   The action of $T_M^\mathrm{in}(0)$  on $\ket{0}_S \otimes \ket{\psi_S}_I \otimes \ket{0}_R^{\otimes M}$ achieves the exact input process from $I$ to $S$.

\begin{figure}[th]
\centering
\includegraphics[width=100truemm]{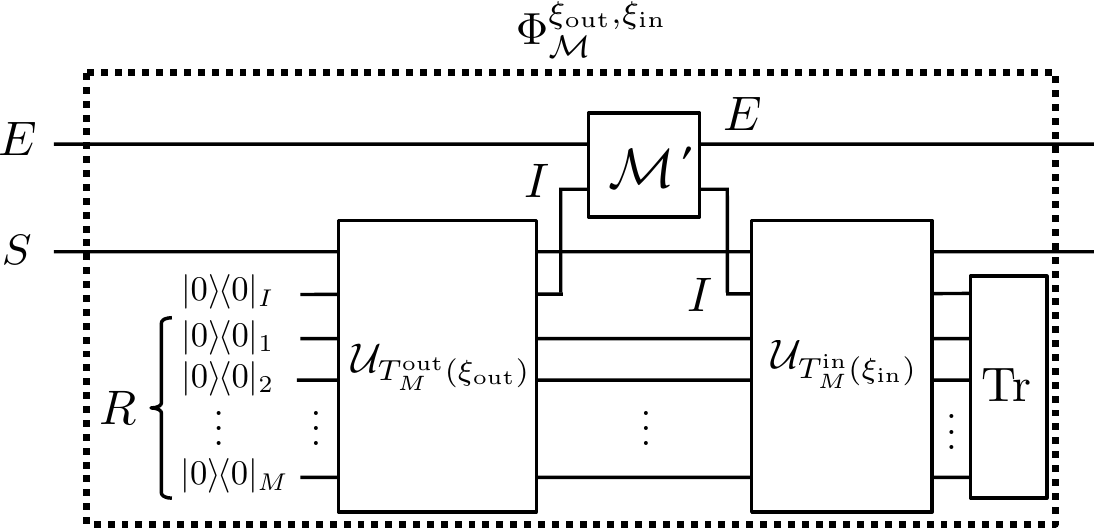}
\caption{A quantum circuit representation of $\Phi _{\mathcal{M}^\prime}^{\xi _\mathrm{out}, \xi _\mathrm{in}}$ defined by  Eq.~(\ref{eq:phioutinM}) for implementing a map $\mathcal{M}'$ on $\mathcal{H}_S \otimes \mathcal{H}_E$.  The initial states of systems in $\mathcal{H}_I \otimes \mathcal{H}_R^M$ are set to be in $\ket{0}_I \otimes \ket{0}_R^{\otimes M}$.
}
\label{fig:outMin}
\end{figure}

We show that we can construct an approximate IO operation by using $T_M^\mathrm{out}(\xi _\mathrm{out})$ and $T_M^\mathrm{in}(\xi _\mathrm{in})$.   In addition to implementing an arbitrary quantum operation $\mathcal{M}$ on $\mathcal{H}_S$, this construction also guarantees to preserve coherence between $S$ and an external system denoted by $E$.   We introduce a quantum operation on $\mathcal{H}_S \otimes \mathcal{H}_E$ denoted by $\mathcal{M}^\prime $.  We define the approximate IO operation as a composite map $\Phi _{\mathcal{M}'}^{\xi _\mathrm{out}, \xi _\mathrm{in}}$ on any linear operator $X$ on $\mathcal{H}_S$ and $\mathcal{H}_ E$ for implementing an arbitrary map $\mathcal{M}^\prime$ by
\begin{align}
\ \Phi _{\mathcal{M}^\prime}^{\xi _\mathrm{out}, \xi _\mathrm{in}}(X)  &:= \mathrm{Tr}_{IR} \Big\{ \Big [ (\mathcal{U}_{T_M^\mathrm{in}(\xi _\mathrm{in})} \otimes \mathcal{I}_{E}) \circ (\mathcal{M}^\prime \otimes \mathcal{I}_{S} \otimes \mathcal{I}_{R})  
  \circ (\mathcal{U}_{T_M^\mathrm{out}(\xi _\mathrm{out})} \otimes \mathcal{I}_{E}) \Big ] \big( X \otimes \outerb{0}{0}_{IR}^{\otimes M+1}\big) \Big \},  \label{eq:phioutinM}
\end{align}
where $\mathrm{Tr}_{IR}$ represents a partial trace operation on system $IR$, $\mathcal{I}_A$ represents an identity map on system $A$ and $\mathcal{U}_{V}$ represent a unitary map corresponding to a unitary operator $V$ on $\mathcal{H}_I \otimes \mathcal{H}_R^M$.  The quantum circuit representation of $\Phi _{\mathcal{M}^\prime}^{\xi _\mathrm{out}, \xi _\mathrm{in}}$ is shown in Fig.~\ref{fig:outMin}.   

The following lemma shows that the difference between the two maps $\Phi _{ \mathcal{M}^\prime}^{\xi _\mathrm{out}, \xi _\mathrm{in}}$ and $\mathcal{M}^\prime$ measured in terms of the diamond norm \cite{W:TQI} is bounded by the parameters $\xi_\mathrm{out}$ and $\xi_\mathrm{in}$.   The norm $\dnm{\bullet}$ denotes the diamond norm for a map $\bullet$.

\begin{Lem} \label{thm:outin}
For any $\mathcal{M}^\prime$ on  $S$ and $E$,  and $0 \leq \xi _\mathrm{out}, \xi _\mathrm{in} \leq 1$, the approximate IO operation  $\Phi _{\mathcal{M}^\prime}^{\xi _\mathrm{out} , \xi _\mathrm{in}}$ satisfies
\begin{align}
\dnm{\Phi _{\mathcal{M}^\prime}^{\xi _\mathrm{out}, \xi _\mathrm{in}} - \mathcal{M}'} \leq \begin{cases}
\; \ 2\sqrt{1-\Xi ^2} \ &(\ \Xi \geq 0\ ) \\
\; \ 2 \ &(\ \Xi < 0 \ )
\end{cases}, \label{eq:boundth1}
\end{align}
where $\Xi := -1 + \sqrt{1-\xi _\mathrm{out}^2} + \sqrt{1-\xi _\mathrm{in}^2}  - \xi _\mathrm{out} \xi _\mathrm{in}.$
\end{Lem}

The proof of Lem.~\ref{thm:outin} is presented in Appendix~\ref{app:thm1}. The lemma shows $\Xi ^2 \approx 1 - (\xi _\mathrm{out} + \xi _\mathrm{in})^2$ for small $\xi _\mathrm{out}$ and $\xi _\mathrm{in}$.  Hence the condition $\dnm{\Phi _{\mathcal{M}^\prime}^{\xi _\mathrm{out}, \xi _\mathrm{in}} - \mathcal{M}^\prime} \leq  \epsilon$ can be achieved for an arbitrary $\epsilon$  by choosing $2 (\xi _\mathrm{out} +\xi _\mathrm{in}) < \epsilon$.  Note that the right hand side of Eq.~(\ref{eq:boundth1}) does not depend on the dimension of $E$.   Lem.~\ref{thm:outin} implies that $T_M^\mathrm{out}(\xi _\mathrm{out})$ and $T_M^\mathrm{in}(\xi _\mathrm{in})$ respectively behave as approximate input and output operations. 
In the following sections, we present concrete constructions of algorithms implementing $T_M^\mathrm{out}(\xi _\mathrm{out})$ and $T_M^\mathrm{in}(\xi _\mathrm{in})$ to achieve the approximate IO operations.   For simplicity, we consider the case of $M = N$, namely, $\mathcal{M} = \mathcal{M}^\prime$ is a unitary operation, but extension for the case of $M \neq N$ is trivial.

\section{The LS algorithm} \label{sec:LS}

\subsection{Quantum circuit representation of the algorithm} \label{ssec:alg1}

The linearly growing size (LS) algorithm achieves the approximate output process of a qubit state from the physical system $S$ to the interface system $I$ using the $N$-qubit register system $R$ in the setting where the allowed operations are in $\mathbb{LU}_{U_\mathrm{int}}$ with $U_\mathrm{int} \in \mathbb{U}^\star$ on  $\mathcal{H}_S \otimes \mathcal{H}_I$ .

\begin{figure}[th]
\centering
\includegraphics[width=110truemm]{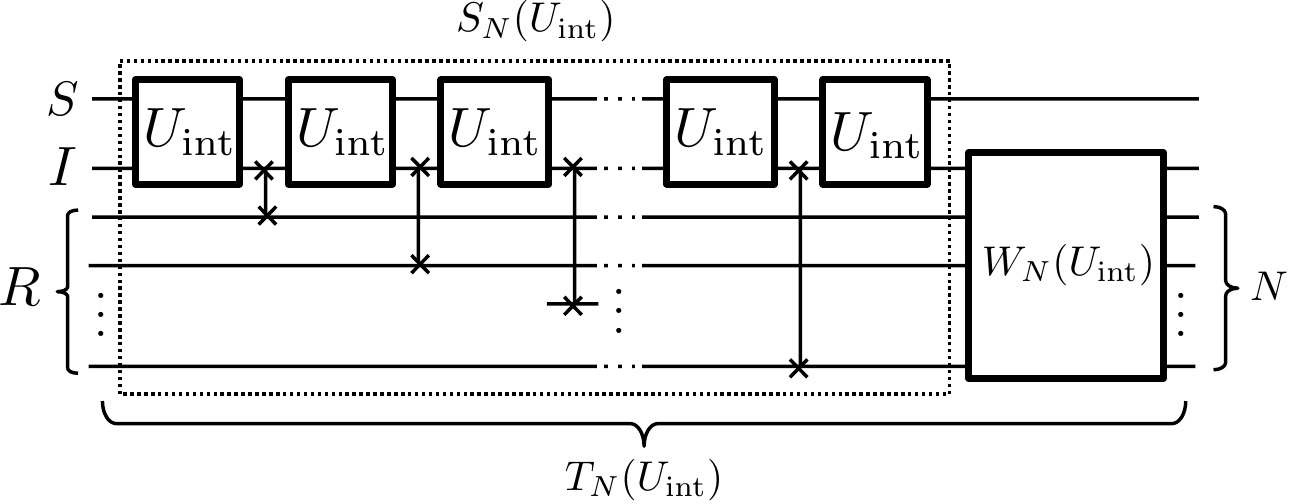}
\caption{A quantum circuit representation of the LS algorithm.   The initial state of system $IR$ is set in $\ket{0}_I \otimes \ket{0}_R^{\otimes N}$ in the feasible basis given by Eq.~(\ref{eq:matU}).  $S_N(U_\mathrm{int})$ consists of alternative iterations of $U_\mathrm{int}$ and swap operations (denoted by a verticle line between quantum wires with ``$\times$" symbols on the wires representing the swapped Hilbert spaces) between the qubits of $I$ and $R_k$ for $k=1,2, \cdots, N$ followed by the final $U_\mathrm{int}$.   $W_N(U_\mathrm{int})$ is a unitary operator on  $\mathcal{H}_I \otimes \mathcal{H}_R^N$ satisfying Eq.~(\ref{eq:wncond}). 
}
\label{fig:alg1}
\end{figure}

The algorithm is the following:
\begin{enumerate}
\item  Set the initial state of the quantum computer (systems $I$ and $R^N$) to be in $\ket{0}_I \otimes \ket{0}_R^{\otimes N}$ where the basis of $I$ is the feasible basis of $U_\mathrm{int}$ satisfying Eq.~(\ref{eq:matU}).  
\item Apply a sequence of operations $T_N (U_\mathrm{int})$ described by a quantum circuit given by Fig.~\ref{fig:alg1}. 
\end{enumerate}
In this algorithm, $T_N(U_\mathrm{int})$ is decomposed into two unitary operators, $S_N(U_\mathrm{int}) \in \mathbb{LU}_{U_\mathrm{int}}^N$ and $W_N(U_\mathrm{int})$ on  $\mathcal{H}_I \otimes \mathcal{H}_R^N$, namely, $T_N(U_\mathrm{int})=(I_S \otimes W_N(U_\mathrm{int})) \cdot  (S_N(U_\mathrm{int}) \otimes I_{R^N})$.  $S_N(U_\mathrm{int})$ transfers quantum information of an arbitrary state from system $S$ to the quantum computer side $I$ and $R^N$, but quantum information is spread over $I$ and $R^N$.  $W_N(U_\mathrm{int})$ retrieve spread quantum information as the state in $I$. $S_N(U_\mathrm{int})$ further consists of alternative iterations of $U_\mathrm{int}$ and swap operations between the qubits of $I$ and $R_k$ for $k=1,2, \cdots, N$ followed by final $U_\mathrm{int}$ as shown in Fig.~\ref{fig:alg1}.   The operation given by $S_N(U_\mathrm{int})$ is also used in the algorithms presented in \cite{DB:FullControl,DB:Protocol}.

We show that $T_N(U_\mathrm{int})$ is an output operation of $T_N^\mathrm{out}(\xi_ \mathrm{out})$ guaranteed to converge to an exact output operation $T_N^\mathrm{out}(0)$ for $N \rightarrow \infty$.   The crux of the LS algorithm is that we always set system $I$ to be in  $\ket{0}_I$ before applying $U_\mathrm{int}$ for each iteration owing to the swap operations. Thus, we only need to consider the action of $U_\mathrm{int}$ on $\ket{\psi_S} \otimes \ket{0}_I = (a_S \ket{0}_S + b_S \ket{1}_S ) \otimes \ket{0}_I$ satisfying $|a_S|^2+|b_S|^2 =1$ for $a_S, b_S \in \mathbb{C}$ , namely, the action of $U_\mathrm{int}$ only on the two basis states $\ket{00}_{SI}$ and  $\ket{10}_{SI}$ as
\begin{align}
\begin{split}
U_\mathrm{int} \ket{00}_{SI} &= \ket{0}_S \otimes (u_{00,00}\ket{0}_I + u_{01,00} \ket{1}_I), 
\\
U_\mathrm{int} \ket{10}_{SI} &= \ket{0}_S \otimes (u_{00,10}\ket{0}_I + u_{01,10} \ket{1}_I) + \ket{1}_S \otimes (u_{10,10}\ket{0}_I + u_{11,10} \ket{1}_I). 
\end{split} \label{eq:00,10}
\end{align}
We define
\begin{align}
\ket{\psi _0} :&= u_{00,00}\ket{0} + u_{01,00} \ket{1},\label{eq:psi0} \\
\ket{\tilde{\psi} _1} :&= u_{00,10}\ket{0} + u_{01,10} \ket{1}, \\
\ket{\tilde{\varphi}} :&= u_{10,10}\ket{0} + u_{11,10} \ket{1}, \label{eq:tphi}
\end{align}
where the tilded states may be unnormalized. Note that $\ket{\psi_0}$ is orthogonal to $\ket{\tilde{\psi} _1}$ due to the unitarily of $U_\mathrm{int}$. 
$\ket{\tilde{\psi} _1}$ and $\ket{\tilde{\varphi}}$ are normalized as
\begin{align}
\ket{\psi _1} := \frac{\ket{\tilde{\psi} _1}}{\alpha}, \ \ \ \ \ket{\varphi} := \frac{\ket{\tilde{\varphi}}}{\beta} \label{eq:psi1+}
\end{align}
by defining
\begin{align}
\alpha  &:= \big\Vert \hspace{-2pt} \ket{\tilde{\psi} _1} \hspace{-2pt} \big\Vert  = \sqrt{|u_{00,10}|^2 + |u_{01,10}|^2},\\ 
\beta  &:= \nm{\ket{\tilde{\varphi}}} = \sqrt{|u_{10,10}|^2 + |u_{11,10}|^2}.
\label{eq:albe}
\end{align}
Note that $\alpha^2+\beta^2=1$ holds due to the unitarity of $U_\mathrm{int}$.
Using the normalized states, the action of the two basis given by Eqs.~(\ref{eq:00,10}) are written by
\begin{align}
U_\mathrm{int} \ket{0}_S \ket{0}_I &= \ket{0}_S \ket{\psi _0}_I,~ \label{eq:002} \\
U_\mathrm{int} \ket{1}_S \ket{0}_I &= \alpha \ket{0}_S \ket{\psi _1}_I + \beta \ket{1}_S \otimes \ket{\varphi}_I. \label{eq:012}
\end{align}

By using Eqs.~(\ref{eq:012}),  the actions of $S_N(U_\mathrm{int})$ on $ \ket{0}_S \ket{\psi _0}^{\otimes N+1}_{IR}$ and $ \ket{1}_S \ket{\psi _0}^{\otimes N+1}_{IR} $ are written by 
\begin{align}
\begin{split}
&S_N(U_\mathrm{int})  \ket{0}_S \otimes \ket{0}_{IR}^{\otimes N+1} = \ket{0}_S \ket{\psi _0}^{\otimes N+1}_{IR}, \\ 
&S_N(U_\mathrm{int})  \ket{1}_S \otimes \ket{0}_{IR}^{\otimes N+1} =\ \alpha \ket{0}_S \otimes \sqb{\sum _{k=1}^{N+1} \beta^{k-1}\ket{\varphi}_{1\dots k-1}^{\otimes k-1} \otimes \ket{\psi _1}_k \otimes \ket{\psi _0}_{k+1\dots N+1}^{\otimes N+1-k}} + \beta^{N+1} \ket{1}_S \otimes \ket{\varphi}^{\otimes N+1}_{IR}, 
\end{split} \label{eq:S00,10}
\end{align}
where $\ket{\bullet}_k$ represents a pure state of $\mathcal{H}_{R_k}$ (the $k$-th qubit of $R$) and $\ket{\bullet }_{N+1}: = \ket{\bullet}_I$. Since $\braket{\psi_0}{\psi_1} = 0$ holds, $\ket{\psi _0}_{IR}^{\otimes N+1}$ and all the states of $\{ \ket{\varphi}_{1\dots k-1}^{\otimes k-1} \otimes \ket{\psi _1}_k \otimes \ket{\psi _0}_{k+1\dots N+1}^{\otimes N+1-k} \}$ are mutually orthogonal.
Therefore, there exists a unitary operator $W_N(U_\mathrm{int})$ on  $\mathcal{H}_{I} \otimes \mathcal{H}_{R}^N$ satisfying
\begin{align}
&W_N(U_\mathrm{int}) \ket{\psi _0}^{\otimes N+1}_{IR} = \ket{0}^{\otimes N+1}_{IR}, \nonumber \\
&W_N(U_\mathrm{int}) \sqb{\sum _{k=1}^{N+1} \beta^{k-1}\ket{\varphi}_{1\dots k-1}^{\otimes k-1} \hspace{-2pt}\otimes \ket{\psi _1}_k \otimes \ket{\psi _0}_{k+1\dots N+1}^{\otimes N+1-k}} =\frac{\sqrt{1-\beta ^{2(N+1)}}}{\alpha } \ket{1}_I \otimes \ket{0}_R^{\otimes N}, \label{eq:wncond}
\end{align}
using the relation $\alpha ^2 + \beta ^2 = 1$.
By applying $W_N(U_\mathrm{int})$ satisfying Eqs.~(\ref{eq:wncond}) after applying $S_N(U_\mathrm{int})$ on an initial state $\ket{\psi _S}_S \otimes \ket{0}_{IR}^{\otimes N+1}$, we obtain
\begin{align}
\ T_N(U_\mathrm{int}) \cdot ( a_S \ket{0}_S + b_S \ket{1}_S) \ket{0}_I \ket{0}_R^{\otimes N} 
=\ \ket{0}_S \otimes \sqb{a_S \ket{0}_I + b_S \sqrt{1-\beta^{2(N+1)}}\ket{1}_I} \otimes \ket{0}^{\otimes N}_R 
+ b_S \beta ^{N+1} \ket{1}_S \otimes \ket{g}_{IR}, \label{eq:LU1}
\end{align}
where $\ket{g}_{IR}$ is a pure state of $\mathcal{H}_I \otimes \mathcal{H}_R^N$.  Since both of $S_N(U_\mathrm{int}),$ $W_N(U_\mathrm{int})$ are elements of  $\mathbb{LU}_{U_\mathrm{int}}^N$, the entire operation $T_N (U_\mathrm{int})$ is also in $\mathbb{LU}_{U_\mathrm{int}}^N$.  Thus $T_N(U_\mathrm{int})$ fulfills the property of $T_N^\mathrm{out} (\beta ^{N+1})$, and when $\beta < 1$, $\beta ^{N+1}$ converges 0 as $N \rightarrow \infty$. Therefore, $T_N(U_\mathrm{int})$ achieves an approximate output operation from the perspective of Lem.~\ref{thm:outin}, if we have a sufficiently large register systems.  

\subsection{Necessary conditions for $S_N(U_\mathrm{int})$} \label{sec:LSdis}

We consider the case when the initial state of the quantum computer is set in a pure state.   In this case, there are two necessary conditions for \emph{general} $S_N(U_\mathrm{int}) \in \mathbb{LU}_{U_\mathrm{int}}^N$ in any output process in the form of $T_N(U_\mathrm{int})=(I_S \otimes W_N(U_\mathrm{int})) \cdot  S_N(U_\mathrm{int})$ satisfying Eq.~(\ref{eq:ToutM}) to guarantee faithful transfer of an arbitrary pure quantum state $\ket{\psi _S}_S$ of system $S$ to system $IR$ (the quantum computer system) in the limit of $N \rightarrow \infty$.   We denote the reduced state of system $S$ after applying $S_N(U_\mathrm{int})$ in the case of $N \rightarrow \infty$ by $\rho _S^\infty$ define by 
\begin{align*}
\rho _S^\infty &:= \lim _{N\to \infty} \Tr _{IR} \big[ S_N(U_\mathrm{int}) \outerb{\psi _S}{\psi _S}_S \otimes \outerb{0}{0}_{IR}^{\otimes N+1}  \times (S_N(U_\mathrm{int}))^\dagger \big] .
\end{align*}

The first condition is that $\rho _S^\infty$ should \emph{not} depend on $\ket{\psi _S}_S$, which is a consequence of the no-cloning theorem of quantum states.  If $\rho _S^\infty$ has some dependence on $\ket{\psi _S}_S$ whereas $\ket{\psi _S}$ is faithfully transferred in system $IR$, some information of $\ket{\psi _S}_S$ can be recovered from $\rho _S^\infty$ in addition to create $\ket{\psi _S}_S$, which violates the no-cloning theorem.    

The second condition is that $\rho _S^\infty$ has to be a \emph{pure state}.  Since the initial states of all $S$, $I$ and $R$ systems are pure states and $S_N(U_\mathrm{int})$ is a unitary operator, the total state of $SIR$ remains in a pure state.  Thus if the reduced state $\rho _S^\infty$ is a mixed state, the total state has to be an entangled state between system $S$ and system $IR$ after applying $S_N(U_\mathrm{int})$.  In such a case, another algorithm to decouple system $S$ from the quantum computer is necessary before performing $\mathcal{M}$ in the quantum computer in our setting.  However, such decoupling does not exist if unitary operations are allowed only on system $IR$.   

We verify that the proposed LS algorithm satisfies these two conditions if and only if $U_\mathrm{int} \in \mathbb{U}^\star $ and the normalization factor $\beta$ defined by Eq.~(\ref{eq:albe}) satisfies $\beta < 1$.  This can be checked as follows.   If $U_\mathrm{int}$ is not in $\mathbb{U}^\star $, the coefficient $a_S$ of the system state $\ket{\psi _S}_S = a_S \ket{0}_S + b_S \ket{1}_S$ spreads to the terms including both $\ket{0}_S$ and $\ket{1}_S$ by applying $U_\mathrm{int}$ and thus $\rho _S^\infty$ becomes a mixed state \cite{TD:PEC}.   If $U_\mathrm{int} \in \mathbb{U}^\star$ but $\beta = 1\ (\alpha = 0)$, $S_N(U_\mathrm{int})$ transforms $(a_S \ket{0}_S + b_S \ket{1}_S) \otimes  \ket{0}_{IR}^{\otimes N+1}$ to $a_S  \ket{0}_S \ket{\psi _0}^{\otimes N+1}_{IR} + b_S \ket{1}_S \otimes \ket{\varphi}^{\otimes N+1}_{IR}$.  The corresponding $\rho _S^\infty$ is a mixed state or remains in $\ket{\psi_S}$ in case $\ket{\psi _0}^{\otimes N+1}_{IR} =\ket{\varphi}^{\otimes N+1}_{IR}$.  On contrary, if $U_\mathrm{int} \in \mathbb{U}^\star$ and $\beta < 1$, $a_S$ does not spread on the terms including  $\ket{1}_S$ while the absolute values of the coefficients of the terms including $\ket{1}_S$ decreases by applying $U_\mathrm{int}$.   Hence the state of $S$ gradually approaches to a pure fixed state $\ket{0}_S$ by $S_N(U_\mathrm{int})$ for $N \rightarrow \infty$, which satisfies the two conditions.  The condition of the convergence of the system state to be a pure fixed state was presented as the \emph{ergodic} condition of the reduced dynamics of system $S$ in the context of full control by locally induced relaxation \cite{DB:FullControl, DB:Protocol}.

\subsection{Implementing $W_N(U_\mathrm{int})$ by applying $U_\mathrm{int}^\dagger$} \label{ssec:Wconst}

In this subsection, we provide a quantum circuit representation of $W_N(U_\mathrm{int})$.  As we have shown in the quantum circuit given by Fig.~\ref{fig:alg1}, $S_N(U_\mathrm{int})$ can be represented \emph{universally} by inserting the swap operations between applications of $U_\mathrm{int}$ irrespective to its classical description as long as $U_\mathrm{int} \in \mathbb{U}^\star$.  Namely, $S_N(U_\mathrm{int})$ is implementable without extracting information about the identity of $U_\mathrm{int}$.  $W_N(U_\mathrm{int})$ is also expressed a function of $U_\mathrm{int}$.   However, it is not known $W_N(U_\mathrm{int})$ can be also represented by iterations of $U_\mathrm{int}$ and a fixed operation.  Instead, we show that $W_N(U_\mathrm{int})$ is implementable by applying $U_\mathrm{int}^\dagger$ by adding an extra qubit in the register system in this subsection.  
Note that it is also unknown that deterministic and universal implementations of $U^\dagger$ by applications of an unknown unitary operation $U$ with a finite usage of $U$.  An algorithm for probabilistic implementations of $U^\dagger$ by applying $U$ for finite times is shown in \cite{Quintino-PRL}, which allows to achieve the probability exponentially close to one in terms of the number of the usage of $U$.

\begin{figure}[th]
\centering
\includegraphics[width=130truemm]{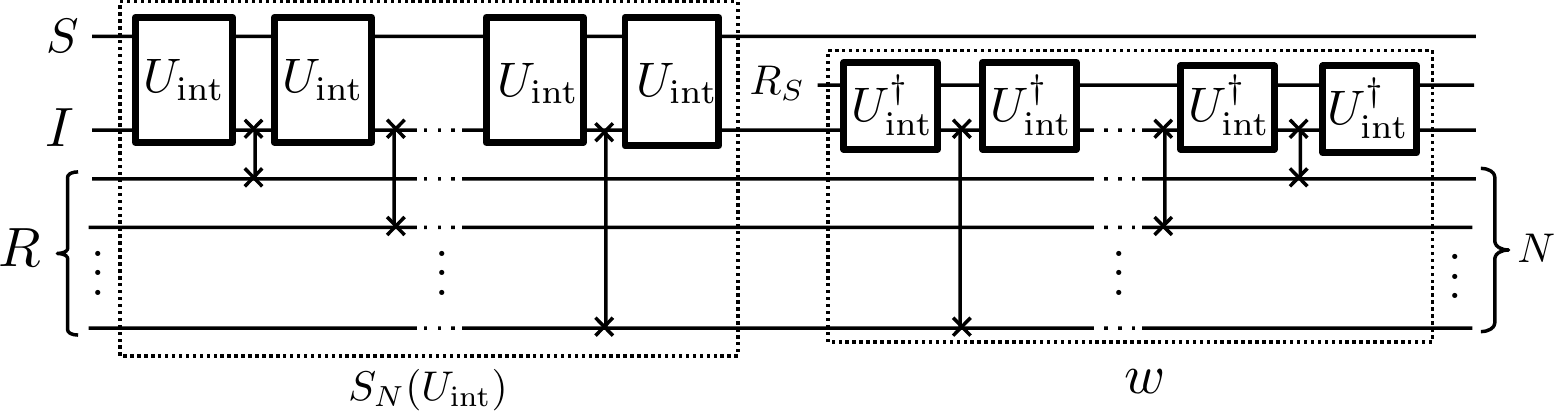}
\caption{A quantum circuit representation of the LS algorithm using $w_N(U_\mathrm{int})$. This circuit consists of only three types of operations; $U_\mathrm{int}$, $U_\mathrm{int}^\dagger$ and swap operations.
}
\label{fig:w+Rs}
\end{figure}

We denote the additional register system as $R_S$ and its Hilbert space as $\mathcal{H}_{R_S}$ in an initial state given by $\ket{0}_{R_S}$.  We define a unitary operator $w_N(U_\mathrm{int}) := (S_N(U_\mathrm{int}))^\dagger$ on  $\mathcal{H}_{R_S} \otimes \mathcal{H}_I \otimes \mathcal{H}_R^N$, which is represented in a part of the quantum circuit shown in Fig.~\ref{fig:w+Rs}.   For any $\ket{\psi _S}_S = a_S\ket{0} + b_S \ket{1}_S$ satisfying $|a_S|^2+|b_S|^2 =1$, a straightforward calculation gives  
\begin{align}
& (I_S \otimes w_N(U_\mathrm{int})) (S_N(U_\mathrm{int}) \otimes I_{R_S}) \ket{\psi _S}_S \ket{0}_{IR}^{\otimes N+1} \ket{0}_{R_S} \nonumber \\
&= \ket{0}_S \otimes \ket{0}_{IR}^{\otimes N+1} \otimes \big( a_S\ket{0}_{R_S} + b_S \sqrt{1-\beta^{2(N+1)}} \ket{1}_{R_S} \big) + \beta^{N+1} \ket{1}_S \otimes \ket{g}_{IRR_S},
\end{align}
using the properties of  $S_N(U_\mathrm{int})$ given by Eqs.~(\ref{eq:S00,10}).  Therefore the action of $W_N(U_\mathrm{int})$ in the LS algorithm can be replaced by  $w_N(U_\mathrm{int})$ followed by a swap operation on $\mathcal{H}_I$ and $\mathcal{H}_{R_S}$ denoted by $U_{\mathrm{swap}(R_S, I)}$ as
\begin{align}
&(I_S \otimes W_N(U_\mathrm{int})) \cdot S_N(U_\mathrm{int}) \ket{\psi _S}_S \ket{0}_{IR}^{\otimes N+1} \ket{0}_{R_S}  \nonumber \\
&=(U_{\mathrm{swap}(R_S, I)} \otimes I_{SR})\otimes (I_S \otimes w_N(U_\mathrm{int}))  \cdot 
(S_N(U_\mathrm{int}) \otimes I_{R_S}) \ket{\psi _S}_S \ket{0}_{IR}^{\otimes N+1} \ket{0}_{R_S}, 
\label{eq:equivW-w}
\end{align}
where $w_N(U_\mathrm{int})$ can be represented \emph{universally} by inserting the swap operations between applications of $U_\mathrm{int}^\dagger$, namely  a function of $U_\mathrm{int}^\dagger$.

The initial states of systems $R_S$, $I$, and $R^N$ have to be set in all $\ket{0}$ in the feasible basis of $U_\mathrm{int}$ satisfying Eq.~(\ref{eq:matU}).  Thus the initial states depend on the classical description of $U_\mathrm{int}$, whereas $S_N(U_\mathrm{int})$ and $w_N(U_\mathrm{int})$ can be given as functions of $U_\mathrm{int}$ and $U_\mathrm{int}^\dagger$ respectively.   The dependence of the initial state of systems $I$, $R_S$, and $R^N$ makes the implementations of $S_N(U_\mathrm{int})$ and $w_N(U_\mathrm{int})$ to be not as quantum supermaps, maps universally transforming a map to another map \cite{ChiribellaSupermaps}.

\section{The CS algorithm} \label{sec:CS}
\subsection{Algorithm} \label{ssec:CSr}

The LS algorithm requires linearly growing register systems $R^N$ to improve the accuracy of the approximation.  We present another algorithm, the CS algorithm, to achieve an approximate output operation requiring a constant-size register, in particular using a single qubit register ($N=1$) in this section.  The guiding principle for constructing the CS algorithm is the necessary conditions for $S_N (U_{\mathrm{int}})$ presented in Sec.~\ref{sec:LSdis}. 

\begin{figure}[th]
\centering
\includegraphics[width=120truemm]{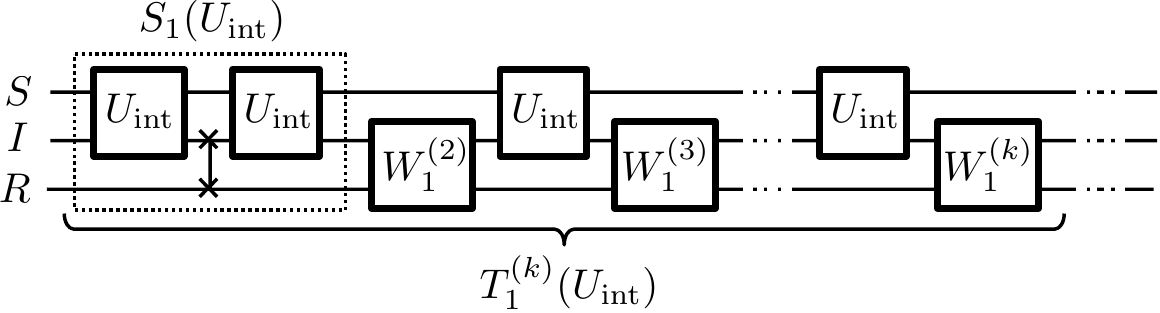}
\caption{A quantum circuit representation of the CS algorithm. The first three steps $S_1(U_\mathrm{int})$ are the same as the LS algorithm. $W_1^{(k)} \ (k=2,3,\cdots )$ are unitary operators on $\mathcal{H}_I \otimes \mathcal{H}_R^1$. 
}
\label{fig:alg2}
\end{figure}

In the CS algorithm, a unitary operator denoted by $W_1^{(k)}$ on $IR$ transforming the state of $I$ in the subspace of $\ket{0}_S$ to $\ket{0}_I$ is inserted after applying $k$th $U_\mathrm{int}$ for $k=2,3,\cdots$ as illustrated in Fig.~\ref{fig:alg2}.  The action of $W_1^{(k)}$ is given by  
\begin{align}
\ket{0}_S \otimes \ket{\tilde{P}}_{IR} + \ket{1}_S \otimes \ket{\tilde{Q}}_{IR} 
\mapsto \ \ket{00}_{SI} \otimes \ket{\tilde{p}}_{R} + \ket{1}_S \otimes \ket{\tilde{q}}_{IR}, \label{eq:roleofW}
\end{align}
where the left hand side state is a general state of $\mathcal{H}_S \otimes \mathcal{H}_I \otimes \mathcal{H}_R^N$ where $\ket{\tilde{P}}_{IR}$, $\ket{\tilde{Q}}_{IR}$, $\ket{\tilde{p}}_{R}$ and $\ket{\tilde{q}}_{IR}$ are appropriate states in the corresponding Hilbert spaces.   The first term of the right hand side state stays in $\ket{0}_S$ by $U_\mathrm{int} \in \mathbb{U}^\star$ similarly to the LS algorithm.  The total gate sequence 
of the CS algorithm is 
\begin{align*}
\ T_1^{(n)} (U_\mathrm{int})  := \ \rb{ \prod _{k=3}^n (I_S \otimes W_1^{(k)}) (U_\mathrm{int} \otimes I_R)} (I_S \otimes W_1^{(2)}) S_1(U_\mathrm{int}), 
\end{align*}
where $S_1(U_\mathrm{int})$ is the case $N=1$ of $S_N(U_\mathrm{int})$ in Fig.~\ref{fig:alg1} and $n$ is the total usage of $U_\mathrm{int}$. By iterating $W_1^{(k)}$ and $U_\mathrm{int}$, we expect that the state of $S$ gradually approaches a pure state $\ket{0}_S$ without using the extra space of $R$.

The detailed analysis of the CS algorithm are shown in App.~\ref{app:thm2}, and we only show how the algorithm works. Suppose an interface unitary $U_\mathrm{int} \in \mathbb{U}^\star$ is given where 
$\{ \ket{i} \otimes \ket{j} \}_{i,j = 0,1}$ is the corresponding feasible basis.  We define $\ket{\psi _0}, \ket{\varphi}$ similarly to the case of the LS algorithm (Eqs.~(\ref{eq:psi0},\ref{eq:tphi},\ref{eq:psi1+})). Then, there exists $T_1^{(n)}(U_\mathrm{int}) \in \mathbb{LU}_{U_\mathrm{int}}^1$ such that
\begin{align}
\ T_1^{(n)}(U_\mathrm{int})  \cdot ( a_S \ket{0}_S + b_S \ket{1}_S) \ket{0}_I \ket{0}_R 
=\ \ket{0}_S \otimes \sqb{a_S \ket{0}_I + b_S \sqrt{1-(\xi ^{(n)})^{2}}\ket{1}_I} \otimes \ket{0}_R + b_S \xi ^{(n)} \ket{1}_S \otimes \ket{g}_{IR}, \label{eq:T1psit}
\end{align}
for any integer $n \geq 2$, which is the total usage of $U_\mathrm{int}$, and complex numbers $a_S , b_S$, where $\ket{g}_{IR}$ is in $\mathcal{H}_{I} \otimes \mathcal{H}_{R}^1$, and $\xi ^{(n)}$ satisfies following;

\begin{itemize}
\item[(A)]~~~~if $|{ \langle \psi_0 | \varphi \rangle} |= 1$,
\begin{align}
\xi ^{(n)} = \beta ^n.
\end{align}
\end{itemize}
Otherwise,
\begin{itemize}
\item[(B-0)]~~~~~~~if $\beta < 1$ and $\omega < 1$,
\begin{align}
\min \{ \beta ^n, \beta ^2\omega ^{n-2} \} \leq \xi ^{(n)} \leq \max \{ \beta ^n, \beta ^2 \omega ^{n-2} \}.  \label{eq:alg2-acc}
\end{align}

\item[(B-1)]~~~~~~~if $\beta < 1$ and $\omega = 1$,
\begin{align}
\beta ^n \leq \xi ^{(n)} \leq \beta ^2 \big( \beta ^2 + (1 - \beta^2) \norm{\psi _0 | \varphi }^2 \big) ^{\frac{n-2}{4}}. \label{eq:B-1}
\end{align}

\item[(B-2)]~~~~~~~if $\beta = 1$ and $\omega < 1$,
\begin{align}
\omega ^{n-2} \leq \xi ^{(n)} \leq \big( \omega ^2 + (1 - \omega^2) \norm{\psi _0| \varphi }^2 \big) ^{\frac{n-2}{4}}, \label{eq:B-2}
\end{align}
\end{itemize}
where
\begin{align}
\beta :=& \sqrt{|u_{10,10}|^2 + |u_{11,10}|^2}, \\
\omega :=& \sqrt{|u_{10,11}|^2 + |u_{11,11}|^2}. \label{eq:omega}
\end{align}

These results imply that $\xi ^{(n)}$ converges to 0 by increasing $n$ by the CS algorithm unless $\beta = 1 \cap (|{ \langle \psi_0 | \varphi \rangle} |= 1 \cup \omega = 1)$, and that the CS algorithm has the same dependence on $n$ as the LS algorithm, where $| \braket{\psi_0} {\varphi} |=1$.   Note that a similar idea of compressing the memory space was introduced in the context of control by locally induced relaxation in \cite{DV:defrag}, where  the necessary memory size was upper bounded by $(\dim \mathcal{H}_S  \dim \mathcal{H}_I)^2 + \dim{\mathcal{H}_I}$. Although their algorithm is applicable for more general cases of $\dim \mathcal{H}_S  \gg 2$, our CS algorithm requires much less memory size of their upper bound.

\subsection{Conditions for interface unitary operators} \label{sec:EIU}

We verify the conditions for interface unitary operators to be satisfiable.  We first define a set of unitary operators $\mathbb{U}^\star$ such that any $U \in \mathbb{U}^\star$ has a local basis $\{ \ket{i}_S \otimes \ket{j}_I \}_{i,j =0,1}$ satisfying
\begin{align}
\bra{1}_S \bra{j}_I U \ket{0}_S \ket{0}_I = 0 \ \text{ for }j=0,1. \label{eq:formcond}
\end{align}
Let $u_{ij,kl} := \bra{i}_S \bra{j}_I U \ket{k} \ket{l}_I$ and $\ket{\psi_0}, \ket{\varphi}, \beta , \omega $ be defined as Eqs.~(\ref{eq:psi0}, -\ref{eq:albe},\ref{eq:omega}).  Then, we refer to $U$ as an {\em exploitable interface unitary operator} unless
\begin{align}
\beta = 1 \cap (\hspace{2pt} |\braket{\psi _0}{ \varphi } |= 1 \ \cup \ \omega = 1). \label{eq:CUcond}
\end{align}

The LS algorithm algorithm implements the approximate output operation $T_N(U_\mathrm{int})$ given by Eq.~(\ref{eq:LU1}) using $N$ qubits register if $\beta < 1$ for $U_\mathrm{int} \in \mathbb{U}^\star$.   On the other hand, the CS algorithm implements the approximate output operation $T_1^{(n)}(U_\mathrm{int})$ in Eq.~(\ref{eq:T1psit}) for an arbitrary positive integer $n$ if $U_\mathrm{int}$ is an exploitable interface unitary.  Therefore, the conditions of $U_\mathrm{int}$ for the CS algorithm is more relaxed than those for the LS algorithm although the CS algorithm requires only one-qubit register.
 
Note that unitary operators satisfying Eqs.~(\ref{eq:formcond}, \ref{eq:CUcond}) are a kind of controlled unitary operators, in addition, all controlled unitary operators fulfilling Eq.~(\ref{eq:formcond}) satisfy Eq.~(\ref{eq:CUcond}).  That is,  when we use a controlled unitary operators as an interface unitary operator, these two algorithms do not work. This result is also obtained by the property of controlled unitary operators in the case of the LS algorithm. To see this, we define a map introduced in \cite{AS:two-p}: 
\begin{align}
\Lambda _{U, \ket{\phi}_I} (X_S) := \Tr _I \big[ U (X_S \otimes \ket{\phi }_I \hspace{-2pt} \bra{\phi }) U^\dagger \big]
\end{align}
for any unitary operator $U$ on $\mathcal{H}_S \otimes \mathcal{H}_I$, $X_S$ on a linear operator on  $\mathcal{H}_S$ and $\ket{\phi}_I$ in $\mathcal{H}_I$.  As pointed out in \cite{AS:two-p}, $\Lambda _{U,\ket{\phi}_I}$ becomes the unital map for any $\ket{\phi}_I$ when $U$ is a local or controlled unitary, i.e., the operator Schmidt number \cite{AN:QD} of $U$ is no more than 2 or, equivalently, the Kraus-Cirac number \cite{BK:OC,AS:two-p} of $U$ ($\# \mathrm{KC}(U)$) is 1 or less.  Consider $U_\mathrm{int}$ to be an interface unitary operator and $\# \mathrm{KC}(U_\mathrm{int}) = 1$, the LS algorithm does not work because the state of $I$ is initialized to a fixed pure state $\ket{0}_I$ before applying $U_\mathrm{int}$.  When $X_S$ is the completely mixed state ($I_S/2$), a unital map $\Lambda _{U_\mathrm{int},\ket{0}_I}$ has no effect on $S$ by definition, which violates the condition of $\rho _S^\infty$.  On the other hand, the CS algorithm contains no initialization for $I$, and it is not obvious whether a controlled unitary operator could be exploitable interface unitary operator. Eqs.~(\ref{eq:formcond}, \ref{eq:CUcond}) implies, however, any exploitable interface unitary operator $U_\mathrm{int}$ fulfills $\# \mathrm{KC}(U_\mathrm{int}) \geq 2$. In other words, all interface interface unitary operators whose KC numbers are less than 2 are inapplicable for the LS/CS algorithms.

\section{Final adjustment of input-output algorithms} \label{sec:rest}

\subsection{Corrections for the effects of $H_S$}

\begin{figure}[th]
\centering
\includegraphics[width=140truemm]{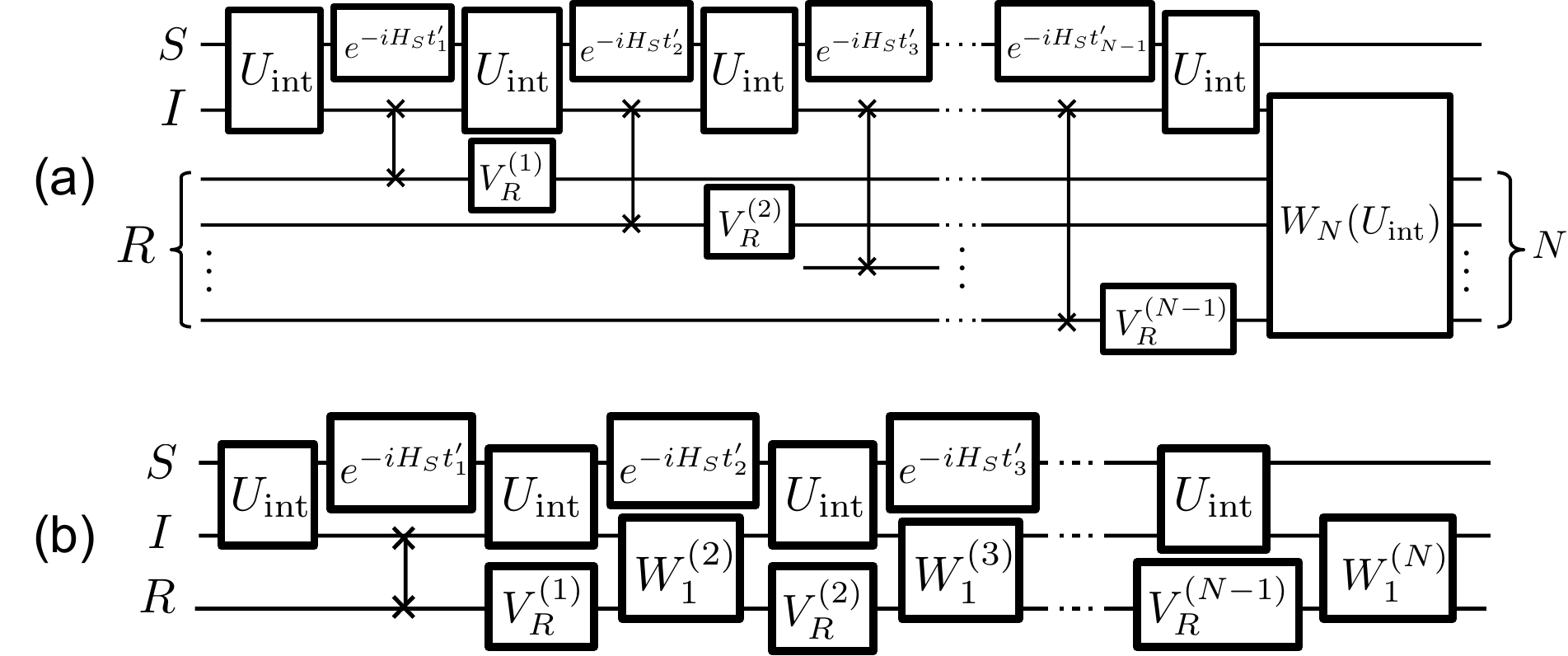}
\caption{Quantum circuit representations of the (a) LS algorithm and (b) CS algorithm including adjustment unitary operators $V_R^{(k)}$.  $V_R^{(k)}$ for $e^{-iH_St_k'}$ are only applied on $R$, hence we can perform them during $U_\mathrm{int}$.
}
\label{fig:Tdag_M}
\end{figure}

So far we only consider the cases $H_S \propto I_S$ or each duration time of $t_\mathrm{off}$ satisfying $e^{-iH_S t_\mathrm{off}} \propto I_S$, namely, we have ignored $H_S$.  However, we will see that the condition $e^{-iH_S t_\mathrm{off}} \propto I_S$ is not necessary for the LS/CS algorithms.  For instance, consider an interface unitary operator $U_\mathrm{int}$ is in $\mathbb{U}^\star$, of which feasible basis of $S$ given by $\{ \ket{i}_S \}_{i=0,1}$ is also the basis diagonalizing $H_S$ as $H_S \ket{j}_S = E_{S, j} \ket{j}_S$ where $E_{S, j}$ is an eigenstate. In this case, we can cancel the effects of $H_S$ in the LS/CS algorithms since $H_S$ does not disturb the weights of $\ket{0}_S$ and $\ket{1}_S$.  That is, the effects of $H_S$ are just phase-shifts for $\ket{1}_S$ such as $e^{-iH_S t'_k}  \ket{0}_S = \ket{0}_S, \ e^{-iH_S t'_k}  \ket{1}_S = e^{-i\theta _k} \ket{1}_S$ where $\theta _k := (E_{S,1} - E_{S,0})t'_k$.  Thus, the state of $S$ at least approaches to $\ket{0}_S$ by the LS/CS algorithms, and in fact, these phase-shifts are always canceled by unitary operations acting on $I$ or $R$. To see this, we note that we can represent all states in $SIR$ on which the phase-shifts act as
\begin{align*}
\ket{\Psi }_{SIR} = \ket{00}_{SI}\ket{\tilde{p}}_R + \ket{10}_{SI}\ket{\tilde{r}}_R + \ket{11}_{SI}\ket{\tilde{s}}_R,
\end{align*}
due to the initializations to $\ket{0}_I$ in $I$ after each SWAP operation in the LS algorithm (thus $\ket{s}_R = 0$) or each $W_1^{(k)}$ in the CS algorithm (see Eq.~(\ref{eq:roleofW})).  By defining $\ket{\tilde{q}}_{IR} := \ket{0}_I\ket{\tilde{r}}_R + \ket{1}_I\ket{\tilde{s}}_R$), the state evolved by the Hamiltonian dynamics of $H_S$ is written by
\begin{align*}
(e^{-iH_S t'_k} \otimes I_I )\ket{\Psi }_{SIR} =\ \ket{00}_{SI}\ket{\tilde{p}}_R + e^{-i\theta _k} ( \ket{10}_{SI}\ket{\tilde{r}}_R + \ket{11}_{SI}\ket{\tilde{s}}_R)
\end{align*}
during the operations on $I$ and $R$ by the phase-shifts.   By Eqs.~(\ref{eq:002},\ref{eq:012}, \ref{eq:112}), we have
\begin{align}
U_\mathrm{int}(e^{-iH_S t'_k} \otimes I_I )\ket{\Psi }_{SIR} 
=\ \ket{0}_S \sqb{\ket{\psi _0}_I \ket{\tilde{p}}_R + e^{-i\theta_k} \ket{\psi _1}_I (\alpha \ket{\tilde{r}}_R + \gamma \ket{\tilde{s}}_R)} + e^{-i\theta _k} \ket{1}_S \ket{\tilde{g}}_{IR}, \label{eq:H_Seff}
\end{align}
where $\ket{\tilde{g}}_{IR}$ is some state of $I$ and $R$.  Since $\ket{\psi _0}$ is orthogonal to $\ket{\psi _1}$, we can cancel the phase $e^{-i\theta _k}$ in the square bracket by a unitary operator on $I$ given by $V_I^{(k)} = \outerb{\psi _0}{\psi _0} + e^{i\theta _k} \outerb{\psi _1}{\psi _1}$.  Next, the sequence of the SWAP or $W_1^{(k+1)}$ in the LS/CS algorithms, respectively, and $e^{-iH_S t'_{k+1}}$ transforms this state into
\begin{align*}
\ket{00}_{SI}\ket{\tilde{p}'}_R + e^{-i(\theta _k+\theta _{k+1})}( \ket{10}_{SI}\ket{\tilde{r}'}_R + \ket{11}_{SI}\ket{\tilde{s}'}_R),
\end{align*}
since the elements with $\ket{0}_S$ are the same as the cases of $H_S \propto I_S$ by the property of $V_I^{(k)}$.  We can cancel the phase $e^{-i(\theta _k+\theta _{k+1})}$ in the second term after $U_\mathrm{int}$ is applied by the above method, thus all phases on $\ket{0}_S$ are adjustable.  In addition, we can apply these operations only applying on $R$.  This is trivial for the LS algorithm because $(V_I^{(k)} \otimes I_R )\mathrm{SWAP} = \mathrm{SWAP} (I_I \otimes V_R^{(k)})$.  In the CS algorithm, the unitary operator $W_1^{(k+1)}$ transforms $\ket{\psi _0}_I \ket{\tilde{p}}_R + e^{-i\theta_k} \ket{\psi _1}_I (\alpha \ket{\tilde{r}}_R + \gamma \ket{\tilde{s}}_R)$ in Eq.~(\ref{eq:H_Seff}) to $\ket{0}_I \ket{\tilde{p}'}_R$, and the orthogonality on $I$ of this vector are completely moved on $R$ (see Eq.~(\ref{eq:Gammak+1})).   We illustrate the final algorithms including the final adjustment of $H_S$ acting on $R$ for the LS and CS algorithms in Fig.~\ref{fig:Tdag_M}.  Therefore, the effects of $H_S$ cause no time loss if the basis $\{ \ket{j}_S \}_{j=0,1}$ diagonalizes $H_S$.

Furthermore, even if $H_S$ is not diagonalized in the $\{\ket{j}_S\}_{j=0,1}$ basis, the LS/CS algorithms work with the above method by choosing the duration time $t'_\mathrm{off}$ such that $e^{-iH_S t'_\mathrm{off}}$ is diagonalized by $\{ \ket{j}_S\}_{j=0,1}$, i.e., $e^{-iH_S t'_\mathrm{off}}$ is just phase-shifts for $\{ \ket{j}_S\}_{j=0,1}$.  Thus, we do not necessarily wait for being $e^{-iH_S t'_\mathrm{off}} \propto I_S$.

\subsection{Designing input operations}

Constructing an approximate input operation is accomplished by the similar way of performing an approximate output operation. In this subsection, we set the number of qubits of the register system back to be $M$.  Lem.~\ref{thm:outin} claims that the adjoint of a trial approximate input operation $T_M^\mathrm{in} (\xi _\mathrm{in})$ has a similar property of $T_M^\mathrm{out}(\xi _\mathrm{out})$ as in Eq.~(\ref{eq:Tin}).  In this sense, we already know how to construct $(T_M^\mathrm{in}(\xi _\mathrm{in}))^\dagger$ in $\mathbb{LU}_{U_\mathrm{int}}^M$ by the LS/CS algorithms. The only difference is we need to perform the adjoint of such operations. To achieve that, we require using $U_\mathrm{int}^\dagger$, but it is not generally allowed under our restrictions except for $U_\mathrm{int} = U_\mathrm{int}^\dagger$. However if $U_\mathrm{int}^\dagger $ is also an exploitable interface unitary operator, the gate sequences $T_M(U_\mathrm{int}^\dagger)$ and $T_1^{(n)}(U_\mathrm{int}^\dagger)$ fulfill the condition for $(T_M^\mathrm{in}(\xi _\mathrm{in}))^\dagger$, and we can implement the adjoint of them in $\mathbb{LU}_{U_\mathrm{int}}^M$.  Fig.~\ref{fig:Tdag_M} illustrates this in the case of the LS algorithm.  Therefore, we can achieve $T_M^\mathrm{in}(\xi _\mathrm{in})$ under our restrictions by $(T_M(U_\mathrm{int}^\dagger))^\dagger$ or $(T_1^{(n)}(U_\mathrm{int}^\dagger))^\dagger$.

\begin{figure}[th]
\centering
\includegraphics[width=110truemm]{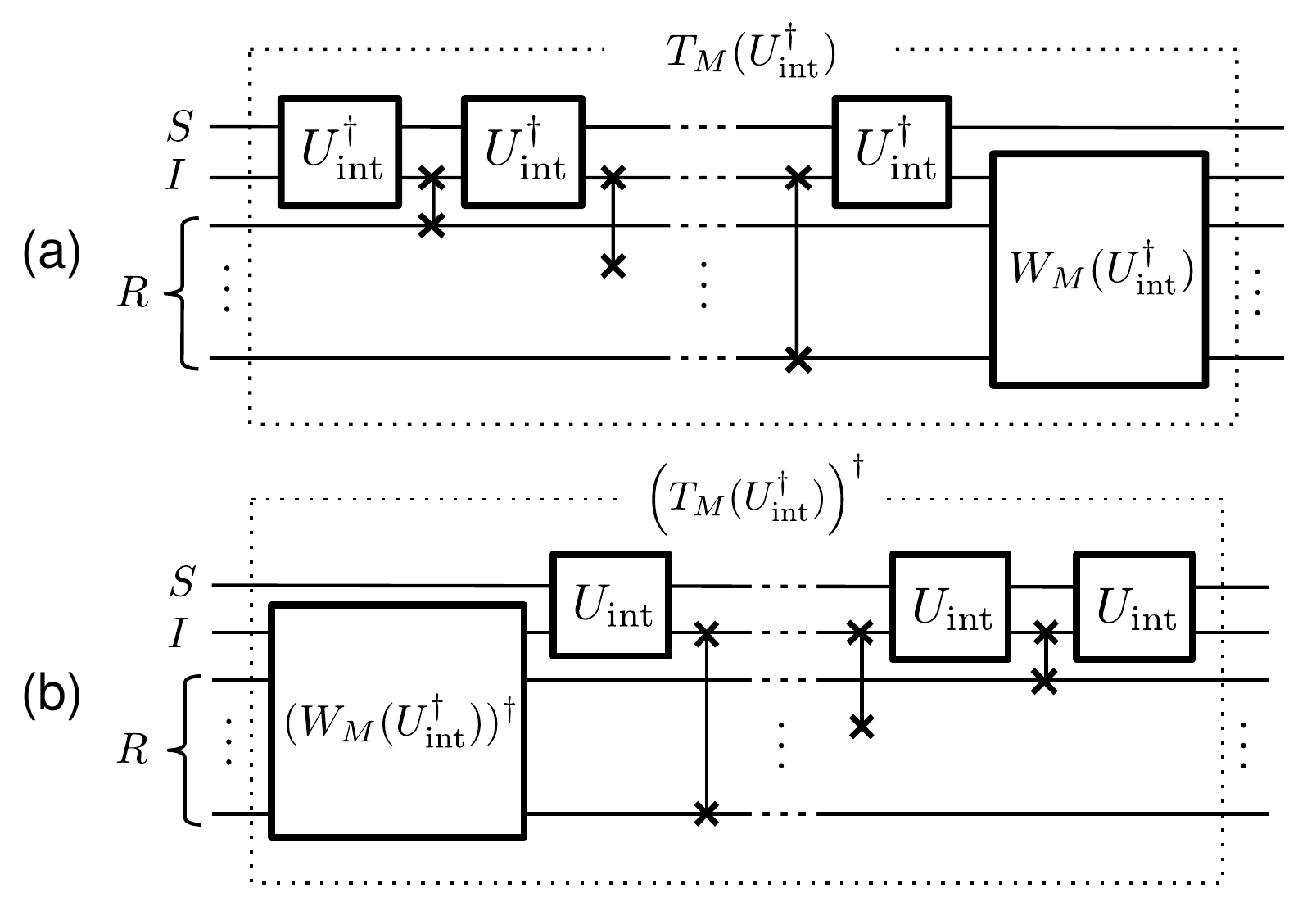}
\caption{(a) A quantum circuit representation of $T_M(U_\mathrm{int}^\dagger)$ according to the LS algorithm with an interface unitary operator $U_\mathrm{int}^\dagger$.  If $U_\mathrm{int}^\dagger$ is exploitable and $\beta < 1$, this operation satisfies the property of $(T_\mathrm{in}^M(\xi ))^\dagger $ in Eq.~(\ref{eq:Tin}).  (b) A quantum circuit representation of the adjoint of (a).  $(T_M(U_\mathrm{int}^\dagger))^\dagger $. This operation is in $\mathbb{LU}_{U_\mathrm{int}}^M$, and can be regarded as input operation $T_M^\mathrm{in} (\xi _\mathrm{in})$. We can use the same technique for the CS algorithm.
}
\label{fig:LSinput}
\end{figure}

There is another restriction to complete input-output algorithms, i,e,. we need to perform $T_M(U_\mathrm{int})$ and $(T_M(U_\mathrm{int}^\dagger)) ^\dagger$ (or, $T_1^{(n)}(U_\mathrm{int})$ and $(T_1^{(n)}(U_\mathrm{int}^\dagger)) ^\dagger$) continuously with keeping their input-output properties.  For this connectivity, $U_\mathrm{int}$ and $U_\mathrm{int}^\dagger$ are restricted to be exploitable interface unitary operators \emph{with the same local basis}, i.e., there must exist a local basis $\{ \ket{i}_S \otimes \ket{j}_I \}_{i,j=0,1} \subset \mathbb{S}(\mathcal{H}_S) \otimes \mathbb{S}(\mathcal{H}_{I})$ such that two matrices $\{ \bra{ij}_{SI} U_\mathrm{int} \ket{kl}_{SI} \}_{ijkl=0,1}$ and $\{ \bra{ij}_{SI} U_\mathrm{int}^\dagger \ket{kl}_{SI} \}_{ijkl=0,1}$ are represented as Eq.~(\ref{eq:matU}) and satisfying Eq.~(\ref{eq:CUcond}). Note that this kind of matrices can be expressed as
\begin{align*}
\begin{bmatrix}
u_{00,00} & u_{00,01} & 0 & 0 \\
u_{01,00} & u_{01,01} & u_{01,10} & u_{01,11} \\
0 & u_{10,01} & u_{10,10} & u_{10,11} \\
0 & u_{11,01} & u_{11,10} & u_{11,11} \\
\end{bmatrix}, 
\end{align*}
but the unitarity of $U_\mathrm{int}$ further restricts the matrices to be given by 
\begin{align}
\begin{bmatrix}
u_{00,00} & 0 & 0 & 0 \\
0 & u_{01,01} & u_{01,10} & u_{01,11} \\
0 & u_{10,01} & u_{10,10} & u_{10,11} \\
0 & u_{11,01} & u_{11,10} & u_{11,11} \\
\end{bmatrix}
\mathrm{~or~}
\begin{bmatrix}
u_{00,00} & u_{00,01} & 0 & 0 \\
u_{01,00} & u_{01,01} & 0 & 0 \\
0 & 0 & u_{10,10} & u_{10,11} \\
0 & 0 & u_{11,10} & u_{11,11} \\
\end{bmatrix}. \label{eq:2possi}
\end{align}
The second matrix of Eqs.~(\ref{eq:2possi}) is a controlled unitary operator,  and not be an exploitable interface unitary operator. On the other hand, the first matrix of Eqs.~(\ref{eq:2possi}) could be an exploitable interface unitary operator for $U_\mathrm{int}$ and $U_\mathrm{int}^\dagger$ unless $u_{01,10} = u_{01,11} = 0$.

\section{Effective interface} \label{sec:Ex}

\begin{figure}[th]
\centering
\includegraphics[width=100truemm]{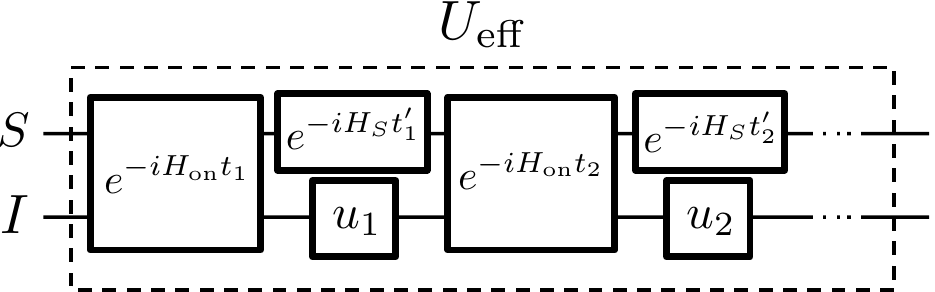}
\caption{
A quantum circuit representation of a set of unitary operators $\mathbb{LU}_{H_S, H_\mathrm{int}}^0$. We perform $e^{-iH_\mathrm{on}t_k}$ and $e^{-iH_S t'_k} \otimes u_k$ for $k=1,2,\dots$ in turn. Each $u_j$ is an arbitrary unitary operator on $\mathcal{H}_I$. We refer to this kind of unitary operators as effective interface unitary operators.
}
\label{fig:Ueff}
\end{figure}

\begin{figure}[th]
\centering
\includegraphics[width=120truemm]{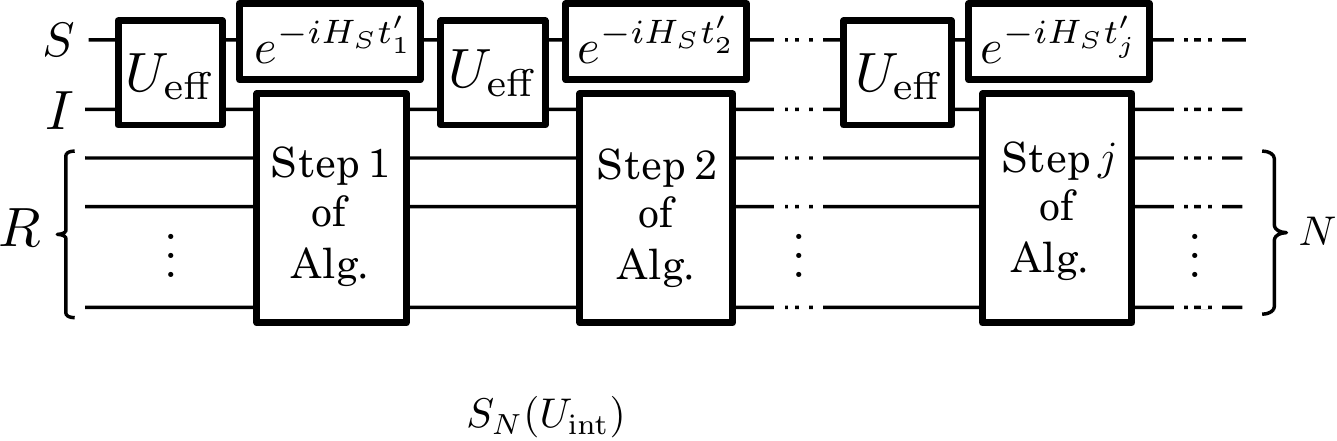}
\caption{
Running our algorithms by using $U_\mathrm{eff} \in \mathbb{LU}_{H_S, H_\mathrm{int}}^0$ instead of a purely time-evolving unitary operator $U_\mathrm{int}(\tau )$. The upper circuit represent an effective interface unitary operator $U_\mathrm{eff}$ which is prepared to be exploitable interface unitary operator by inserting operations on $\mathcal{H}_{I}$.
}
\label{fig:algUeff}
\end{figure}

From all the above results, we can apply arbitrary quantum operations on the system $S$ by only discovering duration $\tau$ such that an interface unitary operator $U_\mathrm{int}(\tau ) = e^{-iH_\mathrm{on}\tau}$ satisfies the conditions of exploitable interface unitary operators.  However, we cannot always find such duration $\tau$ under a given set of Hamiltonians $\{ H_S,\ H_\mathrm{int} \}$.  Nevertheless, we can improve the interface unitary operation by performing unitary operators on $I$, say an \emph{effective interface unitary operator} $U_\mathrm{eff}$, as represented in Fig.~\ref{fig:Ueff}, which are allowed by our constraint.  The CS algorithm works by $U_\mathrm{eff}$ instead of $U_\mathrm{int}$ as depicted in Fig.~\ref{fig:algUeff} when we can construct $U_\mathrm{eff} \in\mathbb{LU}_{H_S, H_\mathrm{int}}^0$ on $\mathcal{H}_S \otimes \mathcal{H}_I$ which is an exploitable interface unitary operator.  In fact, there are cases where we can construct an effective interface unitary operator $U_\mathrm{eff}$ to be exploitable even if $U_\mathrm{int}(\tau )$ is not exploitable for all $\tau$.

In the rest of this section, we will give two sets of Hamiltonians $\{H_S, H_\mathrm{int} \}$ where \#KC$(e^{-iH_\mathrm{on}\tau}) \leq 1$ for all $\tau$.  We will see that one has no possibility to generate an exploitable interface unitary operator by the technique of $U_\mathrm{eff} \in\mathbb{LU}_{H_S, H_\mathrm{int}}^0$, but the other one is possible.

The first example is a bipartite Hamiltonian; 
\begin{align}
H_\mathrm{int}^{(1)} &= r Z_S \otimes Z_I, \\
 H_S &= gZ_S,
\end{align}
where $r , g$ are real numbers.  One can confirm that $\# \mathrm{KC}(e^{-iH_\mathrm{on}^{(1)}t}) \leq 1$ for all time $t$, where $H_\mathrm{on}^{(1)} := r Z_SZ_I + g Z_SI_I$.  $e^{-iH_\mathrm{on}^{(1)}t}$ and $e^{-igZ_St}$ are described by the matrices as
\begin{align*}
&[e^{-iH_\mathrm{on}^{(1)}t}] = 
\begin{bmatrix}
e^{-i(r + g)t} & 0 & 0 & 0 \\
0 & e^{i(r - g)t} & 0 & 0 \\
0 & 0 & e^{i(r + g)t} & 0 \\
0 & 0 & 0 & e^{-i(r - g)t} 
\end{bmatrix}, \\ 
&[e^{-igZ_St} \otimes I_I ] =r
\begin{bmatrix}
e^{-igt} & 0 & 0 & 0 \\
0 & e^{igt} & 0 & 0 \\
0 & 0 & e^{-igt} & 0 \\
0 & 0 & 0 & e^{igt}
\end{bmatrix},
\end{align*}
where we denote the matrix representation of a linear operator $A$ with the Pauli-$Z$ basis by the square bracket $[A]$. Thus, all matrices in $\mathbb{LU}_{H_S, H_\mathrm{int}^{(1)}}^0$ are described as
\begin{align*}
\begin{bmatrix}
A & \mathbf{0} \\
\mathbf{0} & B
\end{bmatrix},
\end{align*}
with the Pauli $Z$-basis where $A$ and $B$ are two-dimensional unitary matrices.  These unitary operators are obviously controlled unitary operators, and we can conclude that both LS and CS algorithms do not work with the above set of Hamiltonians.  

The second example is
\begin{align}
H_\mathrm{int}^{(2)} = r X_S \otimes X_I, \ \ \ \ \ H_S = gZ_S,
\end{align}
which is similar to the first one, but the direction of the interaction Hamiltonian is different.  In this case, we will show that there exists an exploitable interface unitary operator $U_\mathrm{eff} \in \mathbb{LU}_{H_S,H_\mathrm{int}^{(2)}}^0$ (i.e. $\# \mathrm{KC}(U_\mathrm{eff}) \geq 2$) in spite of $\#\mathrm{KC}(e^{-i(r X_SX_I + g Z_SI_I)t}) \leq 1$ for any $t$, i.e., effective interface unitary operators are useful for the LS and CS algorithms.

\begin{figure}[th]
\centering
\includegraphics[width=70truemm]{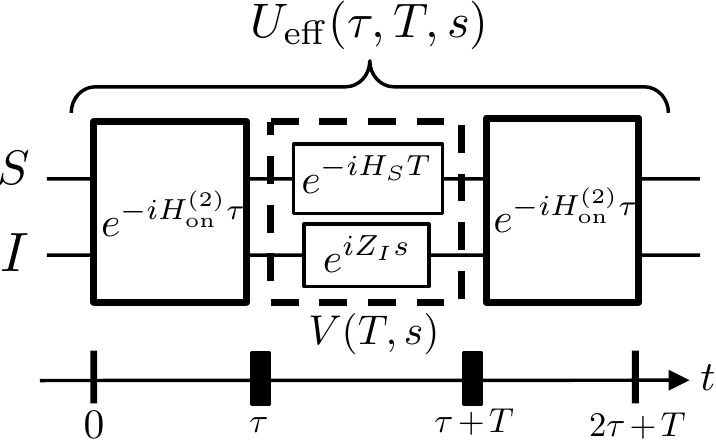}
\caption{
A quantum circuit to construct an exploitable interface unitary operator in $\mathbb{LU}_{gZ_S,r X_SX_I }$.  The total operation is $U_\mathrm{eff} (\tau , T, s )$ and the operation in the dash box is $V(T,s)$.  We switch on the interaction when time $t$ satisfies $0 \leq t < \tau$ or $\tau + T \leq t < 2\tau + T$, and perform $e^{iZ_I s} \in \mathbb{U}(\mathcal{H}_I)$ in $\tau \leq t < \tau + T$.
}
\label{fig:UeffTB}
\end{figure}

Let $H_\mathrm{on}^{(2)} := r X_S X_I + g Z_S I_S$, which is the Hamiltonian for systems $SI$ when the switch of $H_\mathrm{int}^{(2)}$ is on, then $e^{-i[H_\mathrm{on}^{(2)}]t} = [I_{SI}] \cos \bar{\omega} t + \frac{i}{\bar{\omega}}[H_\mathrm{on}^{(2)}] \sin \bar{\omega} t$, where $\bar{\omega} := \sqrt{r ^2 + g^2}$ and $[I_{SI}]$ is an identity matrix for $SI$.  We define $U_\mathrm{eff}(\tau, T, s ) := e^{-iH_\mathrm{on}^{(2)}\tau} V(T, s ) e^{-iH_\mathrm{on}^{(2)}\tau} \in \mathbb{LU}_{H_S,H_\mathrm{int}^{(2)}}^0$ whose quantum circuit is depicted in Fig.~\ref{fig:UeffTB}, where $V(T, s) := e^{-iH_S T} \otimes e^{iZ_I s}$ represents performing a unitary operator $e^{iZ_I s}$ ($s \in \mathbb{R}$) on $I$ while $S$ evolves according to $H_S$ with duration time $T \in (0, \infty)$.  We set $\tau = \tau ^\ast$ such that $\bar{\omega} \tau^\ast = \pi /4$, then by a simple calculation, we obtain
%
\begin{align}
 [U_\mathrm{eff}(\pi/ 4\bar{\omega} , T, s)] 
= \frac{1}{2} [V(T, s )] -\frac{1}{\bar{\omega}}\begin{bmatrix}
g\Theta (gT+s) & & & -ir \Omega (gT + s) \\
& g\Theta (gT-s) & -ir \Omega (gT - s) & \\
& -ir \Omega (gT - s) & g\Theta (-gT+s) & \\
-ir \Omega (gT + s) & & & g\Theta (-gT-s)
\end{bmatrix},
\end{align}
%
where $\Theta (\theta) := ie^{i\theta} + \frac{g}{\bar{\omega}}(e^{i\theta} + \frac{r ^2}{g^2}e^{-i\theta})$ and $\Omega (\theta) := \cos \theta + \frac{g}{\bar{\omega}} \sin \theta$.  There exists $\theta ^\ast$ such that $\Omega (\theta ^\ast) = 0$, i.e.,
\begin{align}
\tan \theta ^\ast = -\frac{\bar{\omega} }{g}.
\end{align}
Therefore, $U_\mathrm{eff}(\pi/ 4\bar{\omega}, T, s )$ becomes an exploitable interface unitary operator when $gT+s = \theta ^\ast + 2m \pi $ and $s \neq n \pi$, where $m,n \in \mathbb{Z}$.  By setting $T=T^\ast := \theta ^\ast /2g$ and $s = s^\ast := \theta ^\ast /2$, the $\beta$ of $U_\mathrm{eff}$ defined in Eq.~(\ref{eq:albe}) is $r/\bar{\omega}$.  When we demand the error less than $\xi _\epsilon$, the usage of $U_\mathrm{eff}$ is bounded by $N^\ast := \lceil \log \xi_\epsilon / \log (r/\bar{\omega}) \rceil$.  Thus, we expect that the total time cost of the LS/CS algorithms is $\mathcal{O}((\tau ^\ast + T^\ast + \Delta T)N^\ast) = \mathcal{O}((\frac{\bar{\omega + g}}{\bar{\omega}g} + \Delta T)\frac{\log \xi_\epsilon}{\log (r/\bar{\omega})})$, where $\Delta T$ is the average time cost to perform the operations on $IR$ (SWAPs or $W_1^{(k)}$) in the LS/CS algorithms.

The difference between the two sets $\{gZ_S,~r Z_S Z_I\}$, $\{gZ_S,~r X_S X_I\}$ are commutativity between $H_S$ and $H_\mathrm{int}$. Since the first set is commutable, nothing induces $S$ to be controlled except for acting $\propto Z_S$. However, the second one is not commutable, thus it can make $S$ controlled by time-evolving $H_S$ and $H_\mathrm{on}$ in turn, which is indicated in Lie-algebraic approach of quantum control theory \cite{SL:CQF,A:IQC,DP:QT,AT:MC,SL:CQI,DB:LocalCon,RH:Localq,RH:Controlq}.

\section{Conclusion} \label{sec:conc}

In this preprint we have considered manipulating a physical system by coupling to a quantum computer where the coupling is described by time-independent Hamiltonian, which can be switched on and off.   For that aim, we have presented two algorithms, the LS and CS algorithms to perform approximate input-output operations under a given interface unitary operator, and showed the interface unitary operators should fulfill conditions to be exploitable in Sec.~\ref{sec:LS} and \ref{sec:CS}.  Our result suggests that the requirements of manipulating a physical system is to discover a recipe of establishing an exploitable interface unitary operator in $\mathbb{LU}_{H_S, H_\mathrm{int}}^0$.  Once we find the recipe, the physical system is manipulatable by the input-output approach.  

The LS algorithm requires a $N$-qubit register system for approximate input-output operations.  The conditions for interface unitary operators are shown to be in $U^\star$ and $\beta < 1$.  In contrast, the CS algorithm requires only a single-qubit register system to perform input-output operations within an arbitrary error.  The conditions for interface unitary operations are shown to be in $U^\star$, but $\beta < 1$ is not necessary and this condition can be relaxed (see (B-0), (B-1) and (B-2) in Sec.~\ref{sec:CS}).  Therefore, the CS algorithm is more widely applicable than the LS one despite the weaker requirement of a constant size register.  The LS algorithm has an advantage that the sequence of inserted gates of the this algorithm is universal, namely, it is independent of the description of the interface unitary operators as shown in Sec.~\ref{ssec:Wconst}, whereas such a universal sequence is not found for the CS algorithm. 

In Sec.~\ref{sec:Ex}, we gave two sets of Hamiltonians which do not generate an exploitable interface unitary operator by time evolution.  We confirmed that one of those has no possibility to construct an exploitable interface unitary operator even if operations on $I$ are included, but that the other has the possibility with that technique.  These results are not surprising in terms of quantum control theory, i.e., input-output operations are realized by the Hamiltonians set $\{ gZ_S , r X_SX_I\}$ and operations on $I$ without $R$.  We do not know a general method realizing an output operation with $U_\mathrm{eff}$.  However, to construct one of exploitable interface unitary operators is generally less demanding than to directly perform an output operation.  

Finally, we should comment that we have only considered a physical system of a 2-dimensional (qubit) system, but we can generalize for arbitrary dimensional systems as considered in \cite{DB:FullControl,DB:Protocol}.  When the physical system $S$ is $(N+1)$-dimensional, one can verify that the matrix form of exploitable interface unitary operations $U_\mathrm{exp}$ on $SI$ can be represented as
\begin{align}
\{ \bra{ij} U_\mathrm{exp} \ket{kl}_{SI} \} = \begin{bmatrix}
u_{00,00} & u_{00,01} & u_{00,10} & \cdots & u_{00,N1} \\
u_{01,00} & u_{01,01} & u_{01,10} & \cdots & u_{01,N1} \\
0 & u_{10,01} & u_{10,10} & \cdots & u_{10,N1} \\
\vdots & \vdots & \ddots & \cdots & \vdots \\
0 & u_{N1,01} & u_{N1,10} & \cdots & u_{N1,N1} \\
\end{bmatrix} \nonumber
\end{align}
with a local basis $\{ \ket{i}_S \} _{i=0}^N \otimes \{ \ket{j}_I \} _{j=0,1}$, and some elements $u_{ij,kl}$ need to be non zero since the final state of $S$ must converge on $\ket{0}_S$ by algorithms. Moreover, if $u_{01,00} = u_{00,01} = \cdots = u_{00,N1} = 0$, we could perform the approximate input-output operations in a sequence by using $U_\mathrm{exp}$.

\section*{Acknowledgment}
This work is supported by MEXT Q-LEAP Grant No. JPMXS0118069605,  JSPS KAKENHI (Grant No.16H01050, 17H01694, 18H04286), and Advanced Leading Graduate Course for Photon Science of Faculty  of Science, The University of Tokyo.



\section{Appendix} 

\emph{Notations:}  For the Hilbert spaces denoted by $\mathcal{H}_1, \mathcal{H}_2$, $\mathbb{L}(\mathcal{H}_1,\mathcal{H}_2)$ and $\mathbb{U}(\mathcal{H}_1,\mathcal{H}_2)$ represent the sets of all linear operators and all unitary operators on $\mathcal{H}_1$ to $\mathcal{H}_2$, respectively. The set of all deterministic quantum operations on a state of $\mathcal{H}_1$ to $\mathcal{H}_2$, which is equivalent to the set of all completely positive trace-preserving maps from $\mathcal{H}_1$ to $\mathcal{H}_2$ is represented by $\mathcal{C}(\mathcal{H}_1, \mathcal{H}_2)$.  For simplicity, $\mathbb{L}(\mathcal{H}_1)$, $\mathbb{U}(\mathcal{H}_1)$ and $\mathcal{C}(\mathcal{H}_1)$ represent the case of $\mathcal{H}_1 = \mathcal{H}_2$.   The set of all pure states in $\mathcal{H}_1$ and the set of all density operators on $\mathcal{H}_1$ are denoted by $\mathbb{S}(\mathcal{H}_1)$ and $\mathbb{D}(\mathcal{H}_1)$, respectively.    

\subsection{Proof of Lemma \ref{thm:outin}} \label{app:thm1}

\begin{figure}[th]
\centering
\includegraphics[width=110truemm]{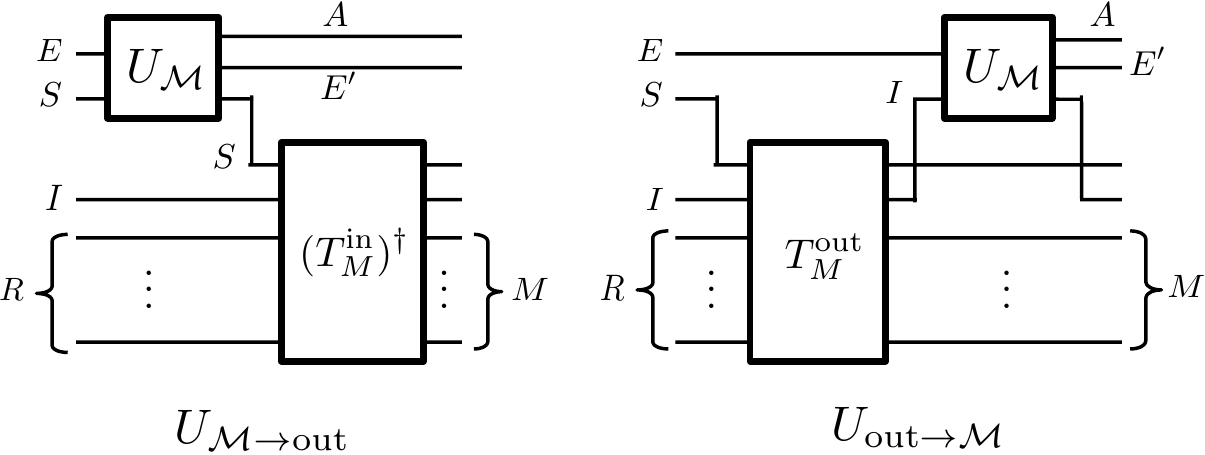}
\caption{
A quantum circuit representations of $U_{\mathcal{M} \to \mathrm{out}}$ and $U_{\mathrm{out} \to \mathcal{M}}$. The left circuit $U_{\mathcal{M} \to \mathrm{out}}$ shows that $U_\mathcal{M}$ is performed at first, then we implement the output operation $(T_\mathcal{M}^\mathrm{out})^\dagger$.  The right circuit $U_{\mathrm{out} \to \mathcal{M}}$ shows that we firstly implement the output operation $T_\mathcal{M}^\mathrm{in}$, then $U_\mathcal{M}$ is performed. The both operations approximate $U_\mathcal{M}$, which acts on $I$ and $E$, instead of $S$ and $E$.}
\label{fig:UMoutoutM}
\end{figure}

\noindent\emph{Proof of } \textbf{Lemma \ref{thm:outin}}\emph{:} The Stinespring representation \cite{W:TQI} of $\mathcal{M} \in \mathcal{C(H}_S\otimes \mathcal{H}_E, \mathcal{H}_S \otimes \mathcal{H}_{E'})$ is
\begin{align}
\mathcal{M}(X) &:= \mathrm{Tr}_A\big[ U_{\mathcal{M}} X U_{\mathcal{M}}^\dagger \big] =: \Psi _{A}^\mathrm{Tr} \circ \mathcal{U}_{U_{\mathcal{M}}}(X)
\end{align}
for any $X \in  \mathbb{L}(\mathcal{H}_S \otimes \mathcal{H}_E)$ by $U_\mathcal{M} \in \mathbb{U}(\mathcal{H}_S \otimes \mathcal{H}_E$, $\mathcal{H}_S \otimes \mathcal{H}_{E'} \otimes \mathcal{H}_A)$ with an appropriate ancillary Hilbert space $\mathcal{H}_A$, and 
%
$
\mathcal{U}_{U_{\mathcal{M}}}(X) := U_{\mathcal{M}} X U_{\mathcal{M}}^\dagger.
$  
We define $\Lambda '_\mathcal{M} \in \mathcal{C}(\mathcal{H}_S \otimes \mathcal{H}_E, \mathcal{H}_S\otimes \mathcal{H}_{E'})$ as
\begin{align}
& \Lambda '_\mathcal{M}(X) 
:=\sqb{ \Phi _{\mathcal{M}}^{\xi_\mathrm{out}, \xi_\mathrm{in}} - \mathcal{M}}(X) \nonumber \\
&=\ \Psi _{IRA}^\mathrm{Tr} \hspace{-1pt} \circ \hspace{-2pt}\Big[ (\mathcal{U}_{T_\mathrm{in}^M} \otimes \mathcal{I}_{E'A}) \circ (\mathcal{U}_{U_{\mathcal{M}}} \otimes \mathcal{I}_{SR}) \circ (\mathcal{U}_{T_\mathrm{out}^M} \otimes \mathcal{I}_{E}) \hspace{70pt} - \mathcal{U}_{U_{\mathcal{M}}} \otimes \mathcal{I}_{IR} \Big] (X \otimes \outerb{0}{0}_{IR}^{\otimes M+1}) \nonumber \\
&=\ \Psi _{IRA}^\mathrm{Tr} \circ \Lambda _\mathcal{M} (X),
\end{align}
where $\Phi _{\mathcal{M}}^{\xi_\mathrm{out}, \xi_\mathrm{in}}$ is in Eq.~(\ref{eq:phioutinM}), and $\Lambda _\mathcal{M}$ 
is
\begin{align}
&\ \Lambda _\mathcal{M} (X) \in \mathcal{C}(\mathcal{H}_S \otimes \mathcal{H}_E, \mathcal{H}_S \otimes \mathcal{H}_I \otimes \mathcal{H}_R \otimes \mathcal{H}_{E'} \otimes \mathcal{H}_A) \nonumber \\
&:=\ \Big[ (\mathcal{U}_{T_\mathrm{in}^M} \otimes \mathcal{I}_{E'A}) \circ (\mathcal{U}_{U_{\mathcal{M}}} \otimes \mathcal{I}_{SR}) \circ (\mathcal{U}_{T_\mathrm{out}^M} \otimes \mathcal{I}_{E}) 
 - \mathcal{U}_{U_{\mathcal{M}}} \otimes \mathcal{I}_{IR}\Big] (X \otimes \outerb{0}{0}_{IR}^{\otimes M+1})
\end{align}
for any $X \in \mathbb{L}(\mathcal{H}_S \otimes \mathcal{H}_E)$.  We omit $\xi _\mathrm{out}$, $\xi _\mathrm{in}$ dependence of $T_\mathrm{out}^M$, $T_\mathrm{in}^M$ in Eqs.~(\ref{eq:Tin}, \ref{eq:ToutM}) for simplicity.  By using this expression, the left hand-side of Eq.~(\ref{eq:boundth1}) satisfies
\begin{align}
\dnm{\Phi _{\mathcal{M}}^{\xi_\mathrm{out}, \xi_\mathrm{in}} - \mathcal{M}} = \dnm{\Psi _{IRA}^\mathrm{Tr} \circ \Lambda _\mathcal{M}}
\leq \dnm{\Lambda _\mathcal{M}}. \label{eq:lambda_Mdnm}
\end{align}
The last inequality is from a property of the diamond norm which monotonically decreases under the action of any map. Since $\Lambda _\mathcal{M}$ consists of isometry maps, $\dnm{\Lambda _\mathcal{M}}$ is equal to $\tnm{\Lambda _\mathcal{M}}$ by Thm.~3.57 of \cite{W:TQI}, where $\tnm{\bullet}$ is the trace norm of maps. Thus, 
\begin{align}
\dnm{\Lambda _\mathcal{M}} = 2\sqrt{1- \min _{\ket{\Psi}_{SE} \in \mathbb{S}(\mathcal{H}_{SE})}\cab{F(\mathcal{M}, \ket{\Psi}_{SE})}^2}, \label{eq:minser}
\end{align}
where $F(\mathcal{M} , \ket{\Psi}_{SE}) $ is defined by
\begin{align}
\ F(\mathcal{M} , \ket{\Psi}_{SE}) 
:= \rb{\bra{\Psi}_{SE} \otimes \bra{0}_{IR}^{\otimes M+1}} U_{\mathrm{out} \to \mathcal{M}}^\dagger 
\times U_{\mathcal{M} \to \mathrm{out}} \rb{\ket{\Psi}_{SE} \otimes \ket{0}_{IR}^{\otimes M+1}},
\end{align}
with
\begin{align}
\begin{split}
U_{\mathcal{M} \to \mathrm{out}} := ((T_\mathrm{in}^M)^\dagger \otimes I_{E'A}) (U_{\mathcal{M}} \otimes I_{IR}),~
U_{\mathrm{out} \to \mathcal{M}} := (U_{\mathcal{M}} \otimes I_{SE})(T_\mathrm{out}^M \otimes I_E).
\end{split}
\end{align}
As depicted in Fig.~\ref{fig:UMoutoutM}, the difference between $U_{\mathcal{M} \to \mathrm{out}}$ and $U_{\mathrm{out} \to \mathcal{M}}$ is the order of the output operations $((T_\mathrm{in}^M)^\dagger $ and $T_\mathrm{out}^M$: see Eqs.~(\ref{eq:Tin}, \ref{eq:ToutM})) and map $\mathcal{U}_\mathcal{M}$.  
$\min F(\mathcal{M}, \ket{\Psi}_{SE})$ is minimizing an inner product of two states
\begin{align}
U_{\mathcal{M} \to \mathrm{out}} &\ket{\Psi}_{SE} \otimes \ket{0}_{IR}^{\otimes M+1},\label{eq:umol} \\
U_{\mathrm{out} \to \mathcal{M}} &\ket{\Psi}_{SE} \otimes \ket{0}_{IR}^{\otimes M+1},\label{eq:uoml}
\end{align}
over $\ket{\Psi}_{SE} \in \mathbb{S}(\mathcal{H}_{S} \otimes \mathcal{H}_{E})$.

Now we represent $\ket{\Psi}_{SE}$ by an orthonormal basis on the physical system $S$, $\{ \ket{0}_S, \ket{1}_S \}$ $\subset \mathbb{S}(\mathcal{H}_S)$ as
\begin{align}
\ket{\Psi}_{SE} := a \ket{0}_S \otimes \ket{e_0}_E + b \ket{1}_S \otimes \ket{e_1}_E \label{eqP2:Pse},
\end{align}
where $|a|^2 + |b|^2 = 1\ (a,b \in \mathbb{C})$ and all elements of $\{ \ket{e_k}_E \}_{k=0,1} \subset \mathbb{S}(\mathcal{H}_E)$ are normalized states on $E$.  In addition, we choose $a_k ,b_k \in \mathbb{C}$ and a set of normalized states $\{ \ket{E_k^j}_{E'A} \}_{j,k=0,1} \subset \mathbb{S}(\mathcal{H}_{E'} \otimes \mathcal{H}_A)$ such that 
\begin{align}
U_{\mathcal{M}} \ket{k}_S \otimes \ket{e_k}_E = a_k \ket{0}_S \otimes \ket{E_k^0}_{E'A} + b_k \ket{1}_S \otimes \ket{E_k^1}_{E'A}, \label{eqP2:UPse}
\end{align}
for $k=0,1$, where $|a_k|^2 + |b_k|^2 = 1$.  Since $U_\mathcal{M}$ is an isometry operator, 
\begin{align}
\bra{0}_S \otimes \bra{e_0}_E U_{\mathcal{M}}^\dagger U_{\mathcal{M}} \ket{1}_S \otimes \ket{e_1}_E = 0
\Rightarrow  a_0^\ast a_1 \braket{E_0^0}{E_1^0} + b_0^\ast b_1 \braket{E_0^1}{E_1^1} = 0. \label{eqP2:restU}
\end{align}
By using the expressions of Eqs.~(\ref{eqP2:Pse}-\ref{eqP2:restU}), 
\begin{align*}
\ U_{\mathcal{M}} \ket{\Psi} _{SE}
= \ket{0}_S \otimes \rb{ aa_0 \ket{E_0^0}_{E'A} + ba_1 \ket{E_1^0}_{E'A}} 
 + \ket{1}_S \otimes \rb{ ab_0 \ket{E_0^1}_{E'A} + bb_1 \ket{E_1^1}_{E'A}}. 
\end{align*}
Then, Eq.~(\ref{eq:umol}) can be described as
\begin{align*}
&\rb{(T_\mathrm{in}^M(\xi _\mathrm{in}))^\dagger \otimes I_{E'A}} \rb{U_\mathcal{M} \ket{\Psi }_{SE} \otimes \ket{0}_{IR}^{\otimes M+1}} \nonumber \\
&= \ket{0}_S \otimes \Big[ \ket{0}_I \otimes \rb{aa_0 \ket{E_0^0}_{E'A} + ba_1 \ket{E_1^0}_{E'A}}  + \eta_\mathrm{in} \ket{1}_I \otimes \rb{ ab_0 \ket{E_0^1}_{E'A} + bb_1 \ket{E_1^1}_{E'A}} \Big] \otimes \ket{0}_R^{\otimes M}  \nonumber \\
&+ \xi_\mathrm{in}\ket{1}_S \otimes \rb{ ab_0 \ket{E_0^1}_{E'A} + bb_1 \ket{E_1^1}_{E'A}} \otimes \ket{g_\mathrm{in}}
\end{align*}
with Eq.~(\ref{eq:Tin}), where $\eta_\mathrm{in} := \sqrt{1-\xi_\mathrm{in}^2}$. On the other hand,  Eq.~(\ref{eq:uoml}) becomes
\begin{align*}
&\rb{U_{\mathcal{M}} \otimes I_{SR}} \rb{T_\mathrm{out}^M(\xi _\mathrm{out})\otimes I_E} \rb{\ket{\Psi}_{SE} \otimes \ket{0}_{IR}^{\otimes M+1}} \nonumber \\
&= \ket{0}_S \otimes \Big[ \ket{0}_I \otimes \Big( aa_0 \ket{E_0^0}_{E'A} + ba_1 \eta _\mathrm{out} \ket{E_1^0}_{E'A}\Big)  \nonumber \\
&+ \ket{1}_I \otimes \Big( ab_0 \ket{E_0^1}_{E'A} + bb_1 \eta  _\mathrm{out} \ket{E_1^1}_{E'A}\Big)\Big] \otimes \ket{0}^{\otimes M}_R   + \xi _\mathrm{out} \ket{1}_S \otimes \big[ (U_{\mathcal{M}} \otimes I_R)\ket{e_0}_{E} \otimes \ket{g_\mathrm{out}}\big].
\end{align*}
where $\eta_\mathrm{out} := \sqrt{1-\xi_\mathrm{out}^2}$.  From the above results,
\begin{align}
&\ F(\mathcal{M}, \ket{\Psi}_{SE}) \nonumber \\
&=\ 1 - (1 - \eta _\mathrm{out}) |b|^2 
+ a^\ast b ( a_0^\ast a_1 \eta_\mathrm{out} \braket{E_0^0 | E_1^0} + b_0^\ast b_1 \eta_\mathrm{out} \braket{E_0^1 | E_1^1})  + a b^\ast ( b_0 b_1^\ast \braket{E_1^1 | E_0^1} + a_0 a_1^\ast \braket{E_1^0 | E_0^0}) \nonumber \\
&- (1 - \eta _\mathrm{in}) (| a b_0 |^2 + \eta_\mathrm{out} | bb_1 |^2  + a^\ast b b_0^\ast b_1 \eta_\mathrm{out} \braket{E_0^1 | E_1^1} + a b^\ast b_0 b_1^\ast \braket{E_1^1 | E_0^1}  )
 + \xi_\mathrm{out} \xi_\mathrm{in} G, \label{eqP2:F1}
\end{align}
where 
\begin{align*}
G:= \sqb{\rb{ ab_0 \bra{E_0^1}_{E'A} + bb_1 \bra{E_1^1}_{E'A}} \otimes \bra{g_\mathrm{in}}} 
\times \sqb{ (U_{\mathcal{M}} \otimes I_R)\ket{e_0}_{E} \otimes \ket{g_\mathrm{out}}},
\end{align*}
and note that $-1 \leq G \leq 1$ because $G$ is constructed from an inner product of two normalized states. By Eq.~(\ref{eqP2:restU}) and the complex conjugate, 
\begin{align*}
 \ F(\mathcal{M}, \ket{\Psi}_{SE}) 
 =\ \xi _\mathrm{out} \xi _\mathrm{in} G +1 -(1 - \eta _\mathrm{out}) |b|^2   - (1 - \eta _\mathrm{in}) (  |ab_0|^2 + \eta _\mathrm{out} |bb_1|^2 + a^\ast b b_0^\ast b_1\eta _\mathrm{out} \braket{E_0^1 | E_1^1} + a b^\ast b_0 b_1^\ast\braket{E_1^1 | E_0^1} ). 
\end{align*}
We investigate the minimum value of $F(\mathcal{M}, \ket{\Psi}_{SE})$ or, equivalently maximize
\begin{align*}
F' (\mathcal{M}, \ket{\Psi}_{SE}) 
:=(1 - \eta _\mathrm{out}) |b|^2 + \rb{1 - \eta _\mathrm{in}} ( |ab_0|^2 + \eta _\mathrm{out} |bb_1|^2 & + a^\ast b b_0^\ast b_1\eta _\mathrm{out} \braket{E_0^1 | E_1^1} + a b^\ast b_0 b_1^\ast \braket{E_1^1 | E_0^1} )
\end{align*}
over $\mathcal{M}$ and $\ket{\Psi} _{SE}$ under $|a|^2 + |b|^2 =1,\ |a_k|^2 + |b_k|^2 =1$ for $k = 0,1$.  The first term of $F' (\mathcal{M}, \ket{\Psi}_{SE})$ is maximum when $|b|^2 = 1$, and the value in the square bracket $[\bullet]$ of the second term always fulfills 
%
$
[\bullet ] \leq \nm{ab_0 \ket{E_0^1}_{E'A} + bb_1 \ket{E_1^1}_{E'A}}^2 = 1
$ 
%
due to $0 \leq \eta _\mathrm{out} \leq 1$.  Thus, 
%
$
F' (\mathcal{M}, \ket{\Psi}_{SE}) \leq 2 - \eta _\mathrm{out} - \eta _\mathrm{in},
$ 
%
where the equality holds if and only if $\xi _\mathrm{out} =0\ (\eta _\mathrm{out} =1)$.  By $F(\mathcal{M}, \ket{\Psi}_{SE}) \geq -\xi _\mathrm{out} \xi _\mathrm{in} + 1 - F'(\mathcal{M}, \ket{\Psi}_{SE})$, 
\begin{align}
F(\mathcal{M}, \ket{\Psi}_{SE}) \geq -1 + \eta _\mathrm{out} + \eta _\mathrm{in} - \xi _\mathrm{out} \xi _\mathrm{in} \label{eqP2:F3}
\end{align}
for any $\mathcal{M} \in \mathcal{C}(\mathcal{H}_S\otimes \mathcal{H}_E, \mathcal{H}_S \otimes \mathcal{H}_{E'})$ and $\ket{\Psi}_{SE} \in \mathbb{S}(\mathcal{H}_{S}\otimes \mathcal{H}_{E})$. Eq.~(\ref{eqP2:F3}) implies
\begin{align}
| F(\mathcal{M}, \ket{\Psi}_{SE}) | \ \geq \ \begin{cases}
\; \ \Xi \ &(\ \Xi \geq 0\ ) \\
\; \ 0 \ &(\ \Xi < 0 \ )
\end{cases}, \label{eq:Xicond}
\end{align}
where
$
\Xi := -1 + \sqrt{1-\xi ^2_\mathrm{out}} + \sqrt{1-\xi ^2_\mathrm{in}} - \xi _\mathrm{out} \xi _\mathrm{in}.
$ 
%
By substituting Eq.~(\ref{eq:Xicond}) for Eq.~(\ref{eq:minser}), we can obtain
\begin{align*}
\dnm{\Lambda _\mathcal{M}} \leq \begin{cases}
\; \ 2\sqrt{1-\Xi ^2} \ &(\ \Xi \geq 0\ ) \\
\; \ 2 \ &(\ \Xi < 0 \ )
\end{cases},
\end{align*}
and by Eq.~(\ref{eq:lambda_Mdnm}) we can verify
\begin{align}
\dnm{\Phi _{\mathcal{M}}^\mathrm{out-in} - \mathcal{M}} &\leq \dnm{\Lambda _\mathcal{M}},  
\leq \begin{cases}
\; \ 2\sqrt{1-\Xi ^2} \ &(\ \Xi \geq 0\ ) \\
\; \ 2 \ &(\ \Xi < 0 \ )
\end{cases}.\label{eqP2:norm3}
\end{align}
\qed

\subsection{The details of the CS algorithm} \label{app:thm2}

To show the result of the CS algorithm, we first prove the following Lemma which corresponds to (A) and (B-0). Then, we consider the cases of (B-1) and (B-2).

\begin{Lem} \label{Thm:pSWAP1}
Let $\mathcal{H}_S, \mathcal{H}_I$ and $\mathcal{H}_{R}^1$ be 2-dimensional Hilbert spaces of the physical system $S$, the interface $I$ and the registers $R$ respectively. Suppose an interface unitary operator $U_\mathrm{int} \in \mathbb{U}^\star$ i.e., by definition, there exists a local basis $\{ \ket{i}_S \otimes \ket{j}_I \}_{ij=0,1} \subset \mathbb{S}(\mathcal{H}_S) \otimes \mathbb{S}(\mathcal{H}_I)$ such that $U_\mathrm{int}$ can be described as Eq.~(\ref{eq:matU}).  We define $\ket{\psi _0}, \ket{\varphi}$ as Eqs.~(\ref{eq:psi0},\ref{eq:psi1+}).  Let a positive integer $N \geq 2$, which is the total usage of $U_\mathrm{int}$, then there exists $T_1^{(N)}(U_\mathrm{int}) \in \mathbb{LU}_{U_\mathrm{int}}^1$ such that
\begin{align}
\ T_1^{(N)}(U_\mathrm{int})  \cdot ( a_S \ket{0}_S + b_S \ket{1}_S) \ket{0}_I \ket{0}_R 
=\ \ket{0}_S \otimes \sqb{a_S \ket{0}_I + b_S \sqrt{1 - (\xi ^{(N)})^2}\ket{1}_I} \otimes \ket{0}_R + b_S \xi ^{(N)} \ket{1}_S \otimes \ket{g}_{IR}, \label{eq:T1psi}
\end{align}
for any $a_S , b_S \in \mathbb{C}$, and $\xi$ satisfies following;

\begin{itemize}
\item[(A)] ~~~~~If $| \left \langle \psi_0 | \varphi \right \rangle | = 1$,
\begin{align}
\xi = \beta ^N.
\end{align}

\item[(B)]~~~~~Otherwise,
\begin{align}
\min \{ \beta ^{N}, \beta ^2\omega ^{N-2} \} \leq \xi \leq \max \{ \beta ^{N}, \beta ^2 \omega ^{N-2} \},
\end{align}
where
\begin{align*}
&\beta := \sqrt{|u_{10,10}|^2 + |u_{11,10}|^2},\\
&\omega := \sqrt{|u_{10,11}|^2 + |u_{11,11}|^2}.
\end{align*}
\end{itemize}
\end{Lem}
\vspace{3mm}

\noindent\emph{Proof} of Lem.~\ref{Thm:pSWAP1}: We define $\ket{\psi _S}, \ket{\tilde{\psi}_1}, \ket{\psi_1}, \ket{\tilde{\varphi}}, \ket{\varphi}$, $\alpha ,\beta , \omega$ as in Eqs.~(\ref{eq:psi0}-\ref{eq:tphi}, \ref{eq:psi1+}, \ref{eq:omega}), and similarly we define
\begin{align}
\begin{split}
&\ket{\tilde{\varphi'}} := u_{10,11}\ket{0} + u_{11,11}\ket{1},\\
&\omega \ket{\varphi '} := \frac{\ket{\tilde{\varphi'}}}{\omega }, \\
&\gamma \ket{\psi _1} := u_{00,11}\ket{0} + u_{01,11} \ket{1}, \\
&\gamma := u_{00,11}\braket{\psi _1 |0} + u_{01,11} \braket{\psi _1|1}.
\end{split}
\label{eq:gamma}
\end{align}
Note that $\nm{\ket{\varphi '}} = 1$ due to  
the unitarity of $U_\mathrm{int}$. Thus, we obtain
\begin{align}
\begin{split}
U_\mathrm{int} \ket{0}_S \ket{0}_I &= \ket{0}_S \ket{\psi _0}_I, \\
U_\mathrm{int} \ket{1}_S \ket{0}_I &= \alpha \ket{0}_S \ket{\psi _1}_I + \beta \ket{1}_S \otimes \ket{\varphi}_I, \\
U_\mathrm{int} \ket{1}_S \ket{1}_I &= \gamma \ket{0}_S \ket{\psi _1}_I + \omega \ket{1}_S \otimes \ket{\varphi '}_I,
\end{split}
\label{eq:112}
\end{align}
where the first two equality are same as Eqs.~(\ref{eq:002}, \ref{eq:012}).

Now we prove this lemma by induction.  
Suppose the state of the total system $SIR$ can be described as
\begin{align}
 \ket{\tilde{\Gamma} ^{(k)}(a,b)}_{SIR} 
:= \ket{00}_{SI} \otimes \rb{a \ket{0}_R + b \eta ^{(k)} \ket{1}_R} + b \rb{\mu ^{(k)}_0 \ket{10}_{SI} \ket{\mu ^{(k)}_0}_R + \mu ^{(k)}_1 \ket{11}_{SI} \ket{\mu ^{(k)}_1}_R} \label{eq:Gammak}
\end{align}
for any $a,b \in \mathbb{C}$, where $\ket{\mu ^{(k)}_0}_R, \ket{\mu ^{(k)}_1}_R \in \mathbb{S}(\mathcal{H}_R^1)$ are mutually orthogonal states in $R$, and $\eta ^{(k)}, \mu ^{(k)}_0$ and $\mu ^{(k)}_1$ are real numbers satisfying
\begin{align}
(\eta ^{(k)})^2 + (\mu ^{(k)}_0)^2 + (\mu ^{(k)}_1)^2 = 1 \label{eq:xm}
\end{align}
to normalize $\ket{\tilde{\Gamma} ^{(k)}(a,b)}_{SIR}$ 
when $|a| ^2 + |b|^2 = 1$. Then, we show {\em (i) there exists operator $V_1^{(k+1)}(U_\mathrm{int}) \in \mathbb{LU}_{U_\mathrm{int}}^1$ which uses $U_\mathrm{int}$ once such that
\begin{align*}
V^{(k+1)}(U_\mathrm{int}) \ket{\tilde{\Gamma} ^{(k)}(a,b)}_{SIR} = \ket{\tilde{\Gamma} ^{(k+1)}(a,b)}_{SIR},
\end{align*}
and $\eta ^{(k)} \leq \eta ^{(k+1)}$.}

To see (i), we divide $V_1^{(k+1)}(U_\mathrm{int})$ into two operations $V^{(k+1)}(U_\mathrm{int}) = (I_S \otimes W_1^{(k+1)}(U_\mathrm{int})) (U_\mathrm{int} \otimes I_R)$, where $W_1^{(k+1)}(U_\mathrm{int}) \in \mathbb{U}(\mathcal{H}_{I} \otimes \mathcal{H}_R^1)$. (We omit the $U_\mathrm{int}$ dependence of $V_1^{(k)}, W_1^{(k)}$ on the rest of this section.) By 
Eqs.~(\ref{eq:112}), 
\begin{align}
 (U_\mathrm{int} \otimes I_R) \ket{\tilde{\Gamma} ^{(k)}(a,b)}_{SIR} 
= \ket{0}_S \otimes \sqb{a \ket{\psi _0}_I \ket{0}_R + b \eta ^{(k+1)} \ket{\eta ^{(k+1)}}_{IR}} + b\ket{1}_S \otimes \ket{\tilde{\nu}^{(k+1)}}_{IR}, \label{eq:UGam}
\end{align}
where we defined $\eta ^{(k+1)}, \ket{\eta ^{(k+1)}}$ where $\Vert \ket{\eta ^{(k+1)}} \Vert = 1$ and $\ket{\tilde{\nu}^{(k+1)}}$ as
\begin{align}
\eta ^{(k+1)} \ket{\eta ^{(k+1)}} 
:=\ \eta ^{(k)} \ket{\psi _0}\ket{1} + \ket{\psi _1} \rb{\alpha \mu ^{(k)}_0 \ket{\mu ^{(k)}_0} + \gamma \mu ^{(k)}_1 \ket{\mu ^{(k)}_1}},\label{eq:xixi}
\end{align}
\vspace{-32pt}

\begin{align}
\eta ^{(k+1)} &:= \sqrt{(\eta ^{(k)})^2 + \cab{\alpha \mu ^{(k)}_0}^2 + \cab{\gamma \mu ^{(k)}_1}^2}, \label{eq:monoxi} \\
\ket{\tilde{\nu}^{(k+1)}} &:= \beta \mu ^{(k)}_0 \ket{\varphi} \ket{\mu ^{(k)}_0} + \omega \mu ^{(k)}_1 \ket{\varphi '} \ket{\mu ^{(k)}_1}, \label{eq:tnu}
\end{align}
and 
$\eta ^{(k+1)} \geq \eta ^{(k)}$. From here, we aim to transform 
\begin{align*}
\ket{0}_S \otimes \ket{\psi _0}_{I} \otimes \ket{0}_R \mapsto \ket{000}_{SIR},~~
\ket{0}_S \otimes \ket{\eta ^{(k+1)}}_{IR} \mapsto \ket{001}_{SIR}
\end{align*}
%
by $W_1^{(k+1)}$. Define projector $P^{(k+1)}_{IR}$ on $\mathcal{H}_{I} \otimes \mathcal{H}_R^1$ as
\begin{align}
P^{(k+1)}_{IR} = \outerb{\psi _0}{\psi _0}_I \otimes \outerb{0}{0}_R + \outerb{\eta ^{(k+1)}}{\eta ^{(k+1)}}_{IR} \label{eq:P}
\end{align}
for $k \geq 2$. By using $P^{(k+1)}_{IR}$, we divide $\ket{\tilde{\nu}^{(k+1)}}_{IR}$ into
\begin{align}
\mu ^{(k+1)}_0 \ket{\nu ^{(k+1)}_0}_{IR} := P^{(k+1)}_{IR} \ket{\tilde{\nu}^{(k+1)}}_{IR},~~ 
\mu ^{(k+1)}_1 \ket{\nu ^{(k+1)}_1}_{IR} := \rb{I_{IR}- P^{(k+1)}_{IR}} \ket{\tilde{\nu}^{(k+1)}}_{IR} ,
\label{eq:Pnu}
\end{align}
where $ \ket{\nu ^{(k+1)}_j}_{IR} \in \mathbb{S}(\mathcal{H}_{I} \otimes \mathcal{H}_R^1)$ described as
\begin{align}
\ket{\nu ^{(k+1)}_0} = \nu_{0,0}^{(k+1)}\ket{0} \ket{\psi _0} + \nu_{0,1}^{(k+1)}\ket{\eta ^{(k+1)}}
\end{align}
by $\nu _{0,j}^{(k+1)} \in \mathbb{C}$.  Note that
\begin{align}
\begin{split}
&|\nu_{0,0}^{(k+1)}|^2 + |\nu_{0,1}^{(k+1)}|^2 = 1, \\
\mu ^{(k+1)}_0 &:= \left \langle \nu^{(k+1)} | P^{(k+1)}_{IR} | \tilde{\nu}^{(k+1)} \right \rangle \geq 0, \\
\mu ^{(k+1)}_1 &:=  \left \langle\nu^{(k+1)} | ( I_{IR} - P^{(k+1)}_{IR}) | \tilde{\nu}^{(k+1)} \right \rangle \geq 0.
\end{split}
\end{align}
Moreover, we define $\ket{\mu ^{(k+1)}_0}_R, \ket{\mu ^{(k+1)}_1}_R \in \mathbb{S}(\mathcal{H}_R^1)$ as
\begin{align}
\ket{\mu^{(k+1)}_0}_R := \nu_{0,0}^{(k+1)} \ket{0}_R + \nu_{0,1}^{(k+1)} \ket{1}_R,
\label{eq:muk+1}
\end{align}

Since $\ket{\psi _0}\ket{0}, \ \ket{\eta ^{(k+1)}}$ and $\ket{\nu_1^{(k+1)}}$ are mutually orthogonal, there exists 
$W_1^{(k+1)} \in \mathbb{U}(\mathcal{H}_{I} \otimes \mathcal{H}_R^1)$ such that
\begin{align}
&W_1^{(k+1)} \ket{\psi _0}_I \ket{0}_R = \ket{0}_I \ket{0}_R,nonumber \\
&W_1^{(k+1)}\ket{\eta ^{(k+1)}}_{IR} = \ket{0}_I \ket{1}_R, \nonumber \\
&W_1^{(k+1)} \ket{\nu_1^{(k+1)}}_{IR} = \ket{1}_I \ket{\mu ^{(k+1)}_1}_R,
\label{eq:W^k+1}
\end{align}
and thus,
\begin{align*}
W_1^{(k+1)} \ket{\nu_0^{(k+1)}}_{IR} 
&=W_1^{(k+1)} \big[ \nu_{0,0}^{(k+1)}\ket{0}_I \ket{\psi _0}_R + \nu_{0,1}^{(k+1)}\ket{\eta ^{(k+1)}}_{IR} \big] \\
&=\ket{0}_I \big( \nu_{0,0}^{(k+1)}\ket{0}_R + \nu_{0,1}^{(k+1)}\ket{1}_R\big) = \ket{0}_I \ket{\mu _0^{(k+1)}}_R.
\end{align*}
By applying $W^{(k+1)}_1$ on Eq.~(\ref{eq:UGam}),
\begin{align}
&(I \otimes W_1^{(k+1)})(U_\mathrm{int} \otimes I) \ket{\tilde{\Gamma }^{(k)}(a,b)}_{SIR}  \nonumber \\
&=\ \ket{00}_{SI} \otimes \rb{a \ket{0}_R + b \eta ^{(k+1)} \ket{1}_R}\nonumber + b \rb{ \mu ^{(k+1)}_0 \ket{10}_{SI} \ket{\mu ^{(k+1)}_0}_R + \mu ^{(k+1)}_1 \ket{11}_{SI}\ket{\mu ^{(k+1)}_1}_R} \nonumber \\
&=\ \ket{\tilde{\Gamma} ^{(k+1)}(a,b)}_{SIR}. \label{eq:Gammak+1}
\end{align}

Therefore, we can verify (i).

Next, we show {\em (ii) we can form $\ket{\tilde{\Gamma }^{(2)}(a,b)}_{SIR}$ from an initial state
\begin{align}
\ket{\psi _S}_S \ket{00}_{IR} = \big( a\ket{0}_S + b\ket{1}_S \big) \ket{00}_{IR}
\end{align}
by performing operations in $\mathbb{LU}_{U_\mathrm{int}}^1$ with two $U_\mathrm{int}$'s.}

The first three steps are $S_1 (U_\mathrm{int})$ in Fig.~\ref{fig:alg1} of $N=1$, i.e.,
\begin{align}
S_1 (U_\mathrm{int}) \ket{\psi _S}_S \ket{00}_{IR}  =  \ket{0}_S \otimes \rb{a\ket{\psi _0}^{\otimes 2}_{IR} + b \alpha \big( \ket{\psi _0}_I \ket{\psi _1}_R + \beta \ket{\psi _1}_I \ket{\varphi}_R \big) } + b\beta ^2\ket{1}_S \ket{\varphi}_{IR}^{\otimes 2}.
\end{align}
In the similar way of Eqs.~(\ref{eq:xixi}, \ref{eq:monoxi}), we define
\begin{align}
\begin{split}
\eta^{(2)} \ket{\eta^{(2)}}_{IR} :=& \alpha \big( \ket{\psi _0}_I\ket{\psi _1}_R + \beta \ket{\psi _1}_I \ket{\varphi}_R \big), \\
\eta^{(2)} :=& \sqrt{\alpha ^2 (1 + \beta ^2)} = \sqrt{1 - \beta ^4} \geq 0,
\end{split}
\end{align}
where we used $\alpha ^2 + \beta ^2 = 1$. By defining $P_{IR}^{(2)}$ as 
\begin{align}
P_{IR}^{(2)} := \outerb{\psi _0}{\psi _0}_{IR}^{\otimes 2} + \outerb{\eta ^{(2)}}{\eta ^{(2)}}_{IR}, \label{eq:P2}
\end{align}
we obtain
\begin{align}
\begin{split}
\mu ^{(2)}_0 \ket{\nu ^{(2)}_0}_{IR} &:= \beta ^2 P_{IR}^{(2)} \ket{\varphi}^{\otimes 2}_{IR}, \\
\ket{\nu ^{(2)}_0}_{IR} &:= \nu_{0,0}^{(2)} \ket{\psi _0}_{IR}^{\otimes 2} + \nu_{0,1}^{(2)}\ket{\eta ^{(2)}}_{IR}, \\
\mu ^{(2)}_1 \ket{\nu ^{(2)}_1}_{IR} &:= \beta ^2(I - P_{IR}^{(2)}) \ket{\varphi}^{\otimes 2}_{IR},\\
\mu ^{(2)}_0 &:= \bra{\varphi}_{IR}^{\otimes 2} P_{IR}^{(2)} \ket{\varphi}_{IR}^{\otimes 2}, \\
\mu ^{(2)}_1 &:= \bra{\varphi}_{IR}^{\otimes 2}(I - P_{IR}^{(2)}) \ket{\varphi}_{IR}^{\otimes 2},\\
\ket{\mu ^{(2)}_0} &:= \nu_{0,0}^{(2)}\ket{0} + \nu_{0,1}^{(2)}\ket{1}, 
\ket{\mu ^{(2)}_1 } \ \ \text{s.t.}\ \ \braket{\mu ^{(2)}_1 | \mu ^{(2)}_0} = 0 
\end{split}\label{eq:mu1}
\end{align}
according to Eqs.~(\ref{eq:Pnu}-\ref{eq:muk+1}) of $k=1$. Since $\ket{\psi _0}\ket{0}$, $\ket{\eta ^{(2)}}$ and $\ket{\nu_1^{(2)}}$ are mutually orthogonal, there exists $W_1^{(2)} \in \mathbb{U}(\mathcal{H}_{I} \otimes \mathcal{H}_R^1)$ such that
\begin{align}
\begin{split}
W_1^{(2)} \ket{\psi _0}_I \ket{0}_R &= \ket{0}_I \ket{0}_R, \\
W_1^{(2)} \ket{\eta ^{(2)}}_{IR} &= \ket{0}_I \ket{1}_R, \\
W_1^{(2)} \ket{\nu_1^{(2)}}_{IR} &= \ket{1}_I \ket{\mu ^{(2)}_1}_R,
\end{split}
\end{align}
and 
%
$
W_1^{(2)} \ket{\nu_0^{(2)}}_{IR} = \ket{0}_I\ket{\mu_0^{(2)}}_R.
$  
%
Thus, we have
\begin{align}
\ (I \otimes W_1^{(2)})S_1(U_\mathrm{int}) \ket{\psi _S}_S \ket{00}_{IR} &= \ \ket{00}_{SI} \otimes \rb{a\ket{0}_R + b \eta ^{(2)} \ket{1}_R} + b \rb{\mu ^{(2)}_0\ket{10}_{SI} \ket{\mu ^{(2)}_0}_R + \mu ^{(2)}_1 \ket{11}_{SI} \ket{\mu ^{(2)}_1}_R} \nonumber \\
&= \ \ket{\tilde{\Gamma }^{(2)}(a,b)}_{SIR}, \label{eq:Gamma1}
\end{align}
which agrees with (ii). Form (i) and (ii), we can conclude that $\eta ^{(k)}$'s are non-decreasing values with increasing $k$.

The rest of proof is on the range of $\eta ^{(k)}$'s. First, we consider the case where $\ket{\varphi} = \ket{\psi _0}$. In this case, we obtain
%
$P_{IR}^{(2)} \ket{\varphi}_{IR}^{\otimes 2} = \ket{\psi _0}_{IR}^{\otimes 2}$ and  
$\ket{\mu ^{(2)}_0} := \ket{0}$ 
%
from Eqs.~(\ref{eq:P2}, \ref{eq:mu1}), thus  we have
\begin{align*}
\ket{\tilde{\Gamma} ^{(2)}(a,b)}_{SIR} = \ket{00}_{SI} \otimes \rb{a\ket{0}_R + b \eta ^{(2)} \ket{1}_R} + b \mu ^{(2)}_0\ket{10}_{SI} \ket{0}_R,
\end{align*}
where $\eta ^{(2)} = \sqrt{1 - \beta ^4},\ \mu _0 ^{(2)} = \beta ^2$. Now we substitute $\ket{\mu _0^{(k)}}_R = \ket{0}_R$ and $\mu _1^{(k)} = 0$ of Eq.~(\ref{eq:Gammak}), then we can see
\begin{align*}
\ V^{(k+1)}_1\ket{\tilde{\Gamma} ^{(k)}(a,b)}_{SIR}  
=\ \ket{00}_{SI}\hspace{-1pt} \otimes \hspace{-1pt} \rb{a \ket{0}_R + b \eta ^{(k+1)} \ket{1}_R}\hspace{-2pt} + b \mu ^{(k+1)}_0 \ket{10}_{SI} \ket{0}_R,
\end{align*}
where 
%
$
1 - (\eta ^{(k+1)})^2 = \beta ^2 (1 - (\eta ^{(k)})^2).
$  
By defining $\xi ^{(k)} := \sqrt{1 - (\eta ^{(k)})^2}$, we can obtain $\xi ^{(k)} = \xi ^{(2)} \beta ^{k-2} = \beta ^{k}$ for $k \geq 2$. Since we use $U_\mathrm{int}$ twice in (ii), the final state of Fig.~\ref{fig:alg2} is
\begin{align}
\ket{\tilde{\Gamma} ^{(N)}(a,b)}_{SIR} &= \ket{00}_{SI} \otimes \rb{a\ket{0}_R + b \eta ^{(N)} \ket{1}_R} + b \mu ^{(N)}_0\ket{10}_{SI} \ket{0}_R, \\
 \eta ^{(N)} &= \sqrt{1 - (\xi ^{(N)})^2} = \sqrt{1 - \beta ^{2N}},
\end{align}
if the total usage of $U_\mathrm{int}$ is $N$.   
We observe that   
$\xi ^{(N)} = \beta ^N$  
when $\ket{\psi _0} = \ket{\varphi} = c \ket{\tilde{\varphi}} \ (c \in \mathbb{C})$ by $T_1^{(N)}(U_\mathrm{int}) \in \mathbb{LU}_{U_\mathrm{int}}$, and we can confirm (A) of Lem.~\ref{Thm:pSWAP1}.

Finally, we consider the case $\ket{\psi _0} \neq \ket{\varphi}$, i.e., the form of $\ket{\tilde{\Gamma} ^{(2)} (a,b)}_{SIR}$ is the same as Eq.~(\ref{eq:Gamma1}), and $\eta ^{(k)}$'s satisfy Eqs. (\ref{eq:xm}, \ref{eq:monoxi}), thus
%
\begin{align}
(\xi ^{(k)})^2 u_\mathrm{min}^2 &\leq  (\xi ^{(k+1)})^2 \leq (\xi ^{(k)})^2 u_\mathrm{max}^2,
\end{align}
%
where $u_\mathrm{min} := \min \{\beta , \omega \},\ u_\mathrm{max} := \max \{\beta , \omega \}$ and we used $\alpha ^2 + \beta ^2 = |\gamma |^2 + \omega ^2 = 1$.  
Then if a total usage of $U_\mathrm{int}$
 is $N$, we can verify
\begin{align}
\beta ^2 u_\mathrm{min}^{N-2} \leq \xi ^{(N)} \leq \beta ^2 u_\mathrm{max}^{N-2}, \label{eq:conc_lem2}
\end{align}
and (B) of Lem.~\ref{Thm:pSWAP1} holds.
\qed
\\

To complete the CS algorithm, we seek the cases of (B-1) and (B-2), i.e., when either $\beta$ or $\omega$ is 1 or, equivalently $\alpha$ or $\gamma$ is 0.   First, we deal with the case of (B-1) ($\omega = 1$). In this case, we need to pay attention to a possibility to be $\mu _0^{(k)} = 0$ of Eq.~(\ref{eq:Gammak}) since $\xi ^{(k+1)} = \xi ^{(k)}$ due to $\omega = 1$ (see Eq.~(\ref{eq:monoxi})).   
Nonetheless, by $U_\mathrm{int} \ket{11} = u_{01,11}\ket{10} + u_{11,11}\ket{11}$, we obtain
\begin{align*}
\ket{\tilde{\Gamma} ^{(k+1)}(a,b)}_{SIR} 
= \ket{00}_{SI} \otimes \rb{a \ket{0}_R + b \eta ^{(k)} \ket{1}_R} + b \big( \mu ^{(k+1)}_0 \ket{10}_{SI}  + \mu ^{(k+1)}_1 \ket{11}_{SI}\big) \ket{\mu ^{(k)}_1}_R,
\end{align*}
where 
\begin{align}
\mu ^{(k+1)}_0 &:= \mu ^{(k)}_1\norm{\psi _0 | \varphi '},\\
\mu ^{(k+1)}_1 &:= \mu ^{(k)}_1 (1-\norm{\psi _0 | \varphi '}^2)^{1/2}.
\label{eq:mub1}
\end{align}
Hence, the element of $\ket{10}_{SI}$ appears in the ($k+1$)th-step, which does not in $k$th-step, and $\eta ^{(k+2)} = [(\eta ^{(k)})^2 + (\alpha \mu ^{(k+1)}_0)^2\ ]^{1/2}$. By using $\alpha^2 + \beta^2 = 1$  
and Eq.~(\ref{eq:xm}), we have
\begin{align}
\xi^{(k+2)} = \xi^{(k)} \sqrt{(1-\norm{\psi_0 | \varphi '}^2) + \beta ^2 \norm{\psi_0 | \varphi '}^2} = \xi^{(k)} \sqrt{\beta^2 + (1-\beta ^2)\norm{\psi_0 | \varphi }^2}. \label{eq:worst1}
\end{align}
The last equality is given by the unitarity of $U_\mathrm{int}$, i.e., $\braket{\varphi}{\varphi '} = 0$ is always satisfied when $\omega = 1$. Eq.~(\ref{eq:worst1}) implies that $\xi^{(k+2)} < \xi^{(k)}$ because the condition (B-1) guarantees that $| \braket{\psi_0}{ \varphi} | < 1$.   
Moreover, it is clear that this decay rate of $\xi$ is the worst one, and the upper bound of the trajectory of $\xi^{(k)}$ is
\begin{align}
\beta ^2 \rb{\beta^2 + (1-\beta ^2)| \braket{\psi_0}{ \varphi} |^2}^\frac{k-2}{4},
\end{align}
which is obtained by the recurrence formula (\ref{eq:worst1}) and $\xi ^{(2)} = \beta ^2$ (i.e., we simulate the case of $\mu_1^{(2j)} = 0$ for all $j = 1, 2, \dots$), and we can verify the statement (B-1).

In the case of (B-2) $(\beta = 1)$,  
we consider the case that $\mu_1^{(k)} = 0$, i.e., 
\begin{align}
\begin{split}
\mu ^{(k+1)}_0 &:= \mu ^{(k)}_0| \braket{\psi_0}{ \varphi} | ,\\ 
\mu ^{(k+1)}_1 &:= \mu ^{(k)}_0 (1-| \braket{\psi_0}{ \varphi} |^2)^{1/2} = \mu ^{(k)}_0 | \braket{\psi_0}{ \varphi} |,
\end{split} \label{eq:mub2}
\end{align}
instead of Eq.~(\ref{eq:mub1}).   By a similar way of (B-1), we obtain
\begin{align}
\xi^{(k+2)} &= \xi^{(k)} \sqrt{\omega^2 + (1-\omega ^2)| \braket{\psi_0}{ \varphi} |^2}. \label{eq:worst2}
\end{align}
From the same reasons of (B-1), the upper bound of the trajectory of $\xi^{(k)}$ is
\begin{align}
\rb{\omega^2 + (1-\omega ^2)| \braket{\psi_0}{ \varphi} |^2}^\frac{k-2}{4},
\end{align}
which is obtained by the recurrence formula (\ref{eq:worst2}) and $\xi ^{(2)} = \beta ^2 = 1$ (i.e., we simulate the case of $\mu_0^{(2j)} = 0$ for all $j = 1, 2, \dots$), and we can verify the statement (B-2).

\end{document}